\documentclass[11pt,a4paper]{article}
\usepackage{jheppub}
\usepackage{bbold}
\usepackage[english]{babel}

\newcommand{\pmat}{\begin{pmatrix}}
\newcommand{\fpmat}{\end{pmatrix}}
\newcommand{\eq}{\begin{equation}}
\newcommand{\feq}{\end{equation}}
\newcommand{\cas}{\begin{cases}}
\newcommand{\fcas}{\end{cases}}

\newcommand{\eqarray}{\begin{eqnarray}}
\newcommand{\feqarray}{\end{eqnarray}}

\newcommand{\be}{\beta}

\def\be{\begin{equation}}
\def\ee{\end{equation}}
\def\bea{\begin{eqnarray}}
\def\eea{\end{eqnarray}}

\title{Spinning-off stringy electro-magnetic memories}
\author{A. Aldi,}
\author{ M. Bianchi,}
\author{M. Firrotta}
\affiliation[a]{Dipartimento di Fisica, Universit\`a di Roma Tor Vergata\\
Via della Ricerca Scientifica 1, 00133, Roma, Italy}
\affiliation[b]{INFN sezione di Roma Tor Vergata \\
Via della Ricerca Scientifica 1, 00133 Roma, Italy}

 \emailAdd{alice.aldi@roma2.infn.it}
 
 \emailAdd{massimo.bianchi@roma2.infn.it}
 
 \emailAdd{mfirrotta@roma2.infn.it}

\abstract{We extend and generalise the string corrections to the EM memory to the Type I superstring including spin effects. Very much as in the simpler bosonic string context, the relevant corrections are non-perturbative in $\alpha'$, slowly decaying (as $1/R$) at large distances and modulated in retarded time $u=t-R$. For  spin $N$ states in the first Regge trajectory they entail a sequence of $N$ derivatives wrt $u$ on the `parent' $N=0$ amplitude. We also briefly discuss how to include loop effects, that broaden and shift the string resonances, and how to modify our analysis for macroscopic semi-classical quasi-BPS coherent states, whose collisions may lead to detectable string memory signals in viable Type I models.}

\makeatletter
\gdef\@fpheader{}
\makeatother

\begin{document}
\maketitle

\section*{{Introduction}}

Recently \cite{AAMBMFshort} we have shown that even at large distances from a string collision the produced EM wave profile receives corrections wrt the one predicted by QED  that can code informations on string resonances. 

The present investigation is a generalisation and extension of the analysis performed in \cite{AAMBMFshort} and, for gravitational waves (GWs),  in \cite{ABFMhet}. In particular we will consider spin effects  and fermions in the case of open unoriented Type I  superstrings. Not too surprisingly we find string corrections to the `spin memory' effects that are expected to take place for photons and gravitons \cite{StromTri} as a result of new soft theorems \cite{NewSoft}, that generalise the well known soft theorems \cite{OldSoft}. 

As a consequence of the leading soft theorems the passing of a wave through a detector leaves a `memory'. In the EM case, the memory consists in a kick (remnant velocity) to the charged particles in the detector\cite{OldEMMemo}. In the gravity case, the memory consists in a permanent displacement of the particles in the detector \cite{OldGravMemo}. The new `spin memory' effect \cite{NewMemo} implies that (massless) particles that counter-rotate with respect to one another acquire a time-delay induced by the flux of angular momentum carried by GWs through the orbits. For charged particles a similar effect takes place. A deep connection between BMSvB symmetry \cite{OldBMS, Barnich, NewBMS}, old and new soft theorems \cite{OldSoft, NewSoft} and old and new Memory Effects \cite{OldEMMemo, NewMemo} has been established in \cite{StromTri}. Experimental tests have been proposed, too \cite{GravMemoExp}.  

String Theory predicts a similar soft behaviour \cite{StringSoft}. In the standard low-energy expansions in $\alpha'$, string corrections are  suppressed and become almost negligible at large distances, where the {EM} and gravitational waves (GWs) seem to retain their `classical' profiles, leaving no obvious room for any string memory\footnote{A different version  of the `string memory effect' was studied by \cite{Afshar1811.07368} that is related to large gauge transformations of the Kalb-Ramond field $B_{MN}$.}. Yet taking into account the infinite tower of string resonances drastically changes the situation. 

In the heterotic string, for instance, a toy model for `BH merger' was studied in \cite{ABFMhet} whereby the large masses of the colliding (BPS) objects leads to a regime with non negligible  corrections to the GWs produced in the process. An analogous phenomenon for {{EM}} waves was shown to take place in the (unoriented) open bosonic string, a variant of Veneziano model with internal dimensions and Wilson lines \cite{MBASsys, MBASWL, BPStor}, that can be more intuitively described in terms of D-branes and $\Omega$-planes \cite{AS,Bianchi:1989du, GPAS, Polch}.

For single photon insertions\footnote{Multi-photon emission would encode non-linear aspects of DBI action, as suggested by Cobi Sonnenschein.} the relevant amplitudes display the familiar soft pole at $\omega =0$, that leads to the known {{EM}} memory effect, plus an infinite tower of simple poles on the real axis, associated to the string resonances.  The latter lead to significant string corrections to the {{EM}} wave profile even at large distance from the collision $R>>L$ that give rise to power series in the variable(s) ${{\zeta}}=\exp(-i{{u}}/2\alpha' n{{\cdot}}p)$, where $u=t-R$ is the retarded time, $n= (1, \vec{x}/R)$ and $p$ is the 4-momentum of one of the charged particles involved in the process. This reveals the non-perturbative nature of the effect in $\alpha'$, that is completely hidden in the usual low-energy expansion.  

At low energy {\it i.e.} for $u >>\alpha' n{{\cdot}}p$, due to destructive interference the effect is washed out. Yet, collisions of cosmic strings or BH mergers may produce a detectable signal for $u\approx \alpha' n{{\cdot}}p$, {\it i.e.} at high energy. The same happens in chiral Type I models \cite{ABPSS}, where more reliable order of magnitude estimates can be made  for phenomenological purposes. 

Very much as in General Relativity and Electro-dynamics, sub-leading terms give rise to `spin memory' effects that we address in the present investigation starting from low but non-zero spin and passing later on to consider higher-spins which are the hallmark of string theory.

Plan of the paper is as follows.

In Section \ref{EMmemostring} we briefly review the EM memory effect both at leading and sub-leading order and describe string corrections thereof.

In Section \ref{Bostringmemo} we generalize the results of \cite{AAMBMFshort} for spin-zero colliding objects in the open unoriented bosonic string to colliding fermions and bosons in the Type I superstring.

In Section \ref{Spin} we present new results for the bosonic string and Type I superstring in the case with higher spins and analyse the modification due to spin effects. 

In Section \ref{Npoint} we discuss higher-point amplitudes and briefly mention higher-loop contributions.

In Section \ref{Coherent}, after estimating the order of magnitude of the colliding objects for the effect to be detectable in phenomenologically viable models such as Type I superstrings with large or warped extra dimensions and TeV scale tension, we discuss how macroscopic semi-classical coherent states can meet the required scales of mass and spin.

In Section \ref{ConcOut} we draw our conclusions and discuss some open issues.

\section{{EM memory and string corrections}}
\label{EMmemostring} 

At large distances from the source $R=|\vec{x}|>\!\!> |\vec{x}'|\approx L$, the retarded potential $A^{{\rm ret}}_{\mu}$ produced by an {{EM}} current ${{J}}_{\mu}$ in Fourier space\footnote{The notation $\widetilde{G}(\omega,\vec{x})$ is for Fourier transform w.r.t. $t$, while $\widehat{G}(\omega,\vec{k})$ is for full transform, also w.r.t. $\vec{x}$.} is given by  
\be
\label{tildeAfromJ}
\widetilde{A}^{\mu}(\omega, \vec{x})=\int d^3x' {e^{{{i}} \omega|\vec{x}{-}\vec{x}'|} \over 4\pi |\vec{x}{-}\vec{x}'|} 
    \widetilde{{J}}^{\mu}(\omega,\vec{x}') 
\approx { e^{{{i}} \omega R} \over  4\pi R} 
\widehat{{J}}^{\mu}(\omega,\vec{k}=\omega\vec{n})\,.
\ee
where $\vec{n}= \vec{x}/R$ is the unit vector in the direction of the observer. In QED the `source' can be any collision, described by a `stripped' amplitude 
\begin{equation}
\widehat{J}_\mu(k;p_j) = {\delta {\cal A}_{n{+}1}(a,k;p_j)\over \delta{a}^\mu(k)} \, .
 \end{equation} 
where $k = \omega(1,\vec{n})$ and $a^\mu(k)$ is the photon polarisation. In the soft limit $k\rightarrow 0$ \cite{OldSoft} 
\be 
{\cal A}^{QED}_{n{+}1}(a,k;p_j) = {g}
\sum_{j=1}^n Q_j {a{\cdot}p_j  + a{\cdot}J_j{\cdot} k \over  k{\cdot}p_j}{\cal A}^{QED}_{n}(p_j)+ ...
\ee
with $g$ the charge quantum and $Q_j$ the charges of the `hard' particles, with momenta $p_j$ and angular momenta $J^{\mu\nu}_j = p_j^\mu \partial^\nu_{p_j} -  p_j^\nu \partial^\mu_{p_j} + S^{\mu\nu}$ that act on the remaining hard amplitude. Absorbing the $\omega$ independent (but non-zero) factor ${\cal A}^{QED}_{n}(p_j)$ into a redefinition of $R$ and with the understanding that $J$ must be replaced by some average spin one has 
\be 
\widetilde{{A}}^{\mu}(\omega, \vec{x}) = {g}  {e^{{{i}} \omega R} \over \omega R}\sum_j Q_j {p_j^{\mu}+J_j^{\mu\nu}k_\nu  \over np_j}  + ... \,,
\ee
where $n^\mu = k^\mu/\omega = (1, \vec{n})$. After Fourier Tranform (FT) in $\omega$, the soft pole at $\omega=0$ produces a constant shift of ${{A}}^{>}_\mu(t,\vec{x})$ at $u=t-R>0$ w.r.t. ${{A}}^{<}_\mu(t,\vec{x})$ at $u=t-R<0$. This is the EM memory effect that, as already said, results in a velocity kick for the charged particles \cite{OldBMS, NewMemo}. More recently the `spin memory effect' due to the sub-leading soft term has been interpreted as a relative de-synchronisation of clocks (time-delay) due to the flux of angular momentum carried by EM waves (or GW or in fact any spinning particle) passing through a loop of counter-rotating (charged) particles \cite{NewMemo, PastPhysRep}. The effect can be measured by Sagnac-like detectors \cite{Sagnac}. In particular Sagnac interferometers are usually employed in laser-based gyroscopic measurements of rotations or for gravitational wave detection such as in LIGO-Virgo \cite{Sagnac}.


In the usual low-energy expansion of String Theory (ST) the retarded potential ${{A}}_{\mu}$ receives corrections in powers of $\alpha' k{{\cdot}}p_j$. The FT in $\omega$ produces terms that vanish faster than $1/R$ at large $R$ and that are thus totally negligible at macroscopic distances. Yet as argued in the introduction, when the masses of the colliding objects are such that $\alpha' k{{\cdot}}p_j\approx 1$, the infinite tower of string resonances produces sizeable corrections even to the leading $1/R$ terms 
\be
\widetilde{{A}}^{\mu}(\omega, \vec{x})= \int d^3x' {e^{{{i}} \omega|\vec{x}{-}\vec{x}'|} \over 4\pi |\vec{x}{-}\vec{x}'|} 
    \widetilde{J}^{\mu}(\omega,\vec{x}'; p_j) 
\approx { e^{{{i}} \omega R} \over  4\pi R} 
{\delta {\cal A}^{ST}_{n{+}1}(a,k;p_j)\over \delta{a}_\mu(k)}\Big\vert_{k^\mu=\omega(1,\vec{n})}\,.
\ee
The corrections are non-perturbative in $\alpha'$, since they form series in  
\be
{{\zeta}}_j=\exp(-i{{u}}/2\alpha' n{{\cdot}}p_j)\ee
that modulates in $u$ the `EM string memory' given by 
\be
{{A}}^{ST}_{\mu}(t, \vec{x}) = {{A}}^{QED}_{\mu}(t, \vec{x}) + \Delta_s{{A}}_{\mu}(t, \vec{x}) \, .
\ee
As shown in \cite{AAMBMFshort}, the unoriented open bosonic string corrections $\Delta_s{{A}}_{\mu} = \theta(u) \Delta_s{{A}}^{>}_{\mu} + \theta(-u) \Delta_s{{A}}^{<}_{\mu}$ satisfy peculiar duality properties $\Delta_s{{A}}^{>}_{\mu} + \Delta_s{{A}}^{<}_{\mu}=0$, that reflect planar duality of disk amplitudes. We will confirm the universal nature of this result for Type I superstrings and spinning states.

For simplicity, we only consider open-string insertions on the boundary.\footnote{Closed-string insertions on the bulk of the disk or the projective plane have been carefully reconsidered in \cite{DiskAmp}.}
At tree level (disk) `color-ordered' amplitudes display soft poles in $kp$ for charged `hard' legs with momentum $p$ adjacent to the soft photon leg  with momentum $k$.  String `massive' poles on the real axis at $2\alpha' kp= -n$ are responsible for the string corrections to the {{EM}} memory.   

Due to quantum loop effects the string resonances acquire a finite width and a mass-shift that are perturbative in $g_s$\footnote{Non-perturbative corrections due to stringy instantons, see for instance \cite{Bianchi:2012ud} can be considered that are expected to be largely suppressed for $g_s{<<}1$.}
and lead to an exponential damping in $u$ of the string memory \cite{ABFMhet}. To a large extent the unstable string resonances play the role of 
quasi-normal modes (QNM's) with ${\rm Im}\omega \neq 0$ of the produced `remnant', an unstable (non-BPS) massive, possibly spinning, state.

\section{  {Type I String Memories}}
\label{Bostringmemo} 

\begin{figure}[h!]
\centering
\includegraphics[scale=0.35]{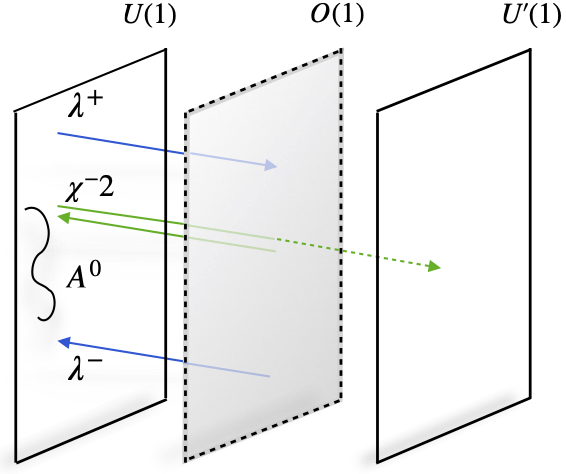}
\caption{Representative picture of the $\Omega3/D3$ setup.}
\label{F1}
\end{figure}

As shown by Green and Schwarz \cite{Green:1984ed}, open and unoriented superstrings in $D=10$ are consistent only for the choice $SO(32)$ of the Chan-Paton group\footnote{For unoriented open and closed bosonic strings with $\Omega$25-plane, dilaton tadpole cancellation  requires $2^{13}$ D25-branes, leading to the gauge group  
$SO(8192)$ \cite{MBAS8192, Alt8192}.}. In lower dimensions, as originally shown in \cite{MBASsys, MBASWL, BPStor} and later on refined in \cite{Witten:1997bs, Bianchi:1997rf}, the gauge group can be broken and the rank can be reduced. Recently new constraints from co-bordism have been identified 
\cite{Montero:2020icj}. Wilson lines in the original D9-brane description are T-dual to brane displacements. Here we work in a local setting with D3-branes and $\Omega$3-planes, equivalent to D9-branes with Wilson lines \cite{MBASsys, MBASWL, BPStor} and ignore global consistency conditions that can be fulfilled by adding well-separated D-branes and/or (discrete) fluxes.  More specifically we consider a 4-dimensional Type I-like configuration, shown in Fig.~\ref{F1}, with one D3-brane on top of an $\widetilde{\Omega{3}}^{-}$-plane, giving rise to an $O(1)$ gauge group, and one D3-brane parallel to but separated from the $\widetilde{\Omega{3}}^{-}$-plane, giving rise to a $U(1)$ gauge group. 

In order to expose the string corrections to the EM memory effect, we consider a 4-point amplitude with a single massless $U(1)$ photon $a_\mu$, two charged (massive) gaugini $\lambda^{(+1)}$, with ends on the $U(1)$ D3-brane and on the $O(1)$ D3-brane, mass $\alpha' M_{\pm 1}^2 = {{\delta}}^2$ (where ${{\delta}}^2 = d^2/\alpha'$)   and one doubly-charged scalar, from the $U(1)$ D3-brane to its image, 
with mass $\alpha' M_{\pm 2}^2 = 4{{\delta}}^2$.  
 
The vertex operators for the $U(1)$ gauge boson $A_\mu$ in the 0-picture reads as usual
\be
\mathcal{V}^{(0)}_{A} = a{\cdot} (i{\partial} X+ 2\alpha'\Psi k{\cdot}\Psi)  e^{ik{\cdot} X} 
\ee 
with $k^2=0$ and $k{\cdot}a(k)=0$. For the singly-charged gaugini in the canonical $-1/2$ picture one has  
\be
\mathcal{V}_\lambda= \sqrt{2\alpha'} \lambda^A_\alpha S^{\alpha} \Sigma_A e^{-\varphi/2} e^{iK{\cdot} X}\ee
with $K=(p^\mu; \pm\vec{d}/\alpha')$, $S^{\alpha}$ ($\alpha =1,2$) a chiral $SO(1,3)$ spin field and $\Sigma_A$ ($A =1,...4$) a chiral $SO(6)$ spin field. For the doubly-charged BPS scalar $\chi^{-2}$ in the canonical $-1$ picture one has 
\be
\mathcal{V}_\chi= \sqrt{2\alpha'} \chi_i \Psi^i  e^{-\varphi} e^{iK{\cdot} X}\ee
 with $K=(p^\mu; \pm 2\vec{d}/\alpha')$ and $i=1,...6$ with $\vec{d}{\cdot}\vec{\chi}=0$. 
 
In addition to the `minimal' couplings of $\lambda^{\pm}$ and $\chi^{\mp2}$ to the photon, 
one has the gauge-invariant Yukawa couplings  
\be
{\cal A}_{\lambda^{\pm}\lambda^{\pm}\chi^{\mp2}}={g_{op}\over \sqrt{2\alpha'}}
\varepsilon^{\alpha\beta} \chi_{AB}(3) \lambda^A_\alpha(1) \lambda^B_\beta(2)\ee 
where $\chi_{AB} = \Gamma^i_{AB} \chi_i$ and $g_{op}{=}\sqrt{g_s}$ is the open string coupling.

This Yukawa coupling is the dimensional reduction of the minimal coupling in $D=10$
\be
{\cal A}_{\Lambda \Lambda A} = g_{op} f_{rst} A_M \Gamma^M_{(ab)} \Lambda^a(1)  \Lambda^b(2)
\ee
which is zero for collinear momenta $K_1\sim K_2\sim K_3$ since 16-components commuting chiral spinors in $D=10$ with collinear momenta $u^a(K_1) \sim u^a(K_2)$ can be chosen to satisfy
$u^a(K)\Gamma^M_{ab} u^b(K)= K^M$ and $A_M K^M=0$.

A non-vanishing 3-point amplitude requires one non-BPS state at least. However, replacing $\chi^{\pm2}$ with 
$H^{\pm2}$, at the first massive level, whose emission vertex reads 
\be
\mathcal{V}_H= H_{ij} \Psi^i \partial X^j e^{-\varphi} e^{iK{\cdot} X}\ee
 with $\alpha' K^2= -1$ {\it i.e.}  $\alpha' m_H^2= 1 + 4 {|\vec{d}|^2\over \alpha'} $ and $\vec{d}^iH_{ij}=0= \delta^{ij}H_{ij}$ would not work! The relevant 3-point amplitude  
 \be
{\cal A}_{\lambda^{\pm}\lambda^{\pm}H^{\mp2}}={g_{op}\over \sqrt{2\alpha'}}
\varepsilon^{\alpha\beta} \lambda^A_\alpha(1) \Gamma^i_{AB}  \lambda^B_\beta(2)H_{ij}(3) (d^j_1-d^j_2)\ee 
vanishes since $\vec{d}_1$, $\vec{d}_2$ and $\vec{d}_3$ are collinear ($\vec{d}_1=\vec{d}_2=\vec{d}$ and $\vec{d}_3= - \vec{d}_1-\vec{d}_2=-2\vec{d}$), so that $\vec{d}_1-\vec{d}_2=0$. The simplest way around, at the same mass level, is to consider 
\be
\mathcal{V}_C= {\sqrt{2\alpha'}}C_{ijk} \Psi^i \Psi^j \Psi^k e^{-\varphi} e^{iK{\cdot} X}\ee
with $\alpha' m_C^2= 1 + 4 {|\vec{d}|^2\over \alpha'} $ and ${d}^iC_{ijk}=0$.
The relevant 3-point amplitude is 
 \be
 \label{nonzero3pt}
{\cal A}_{\lambda^{\pm}\lambda^{\pm}C^{\mp2}}={g_{op}\over \sqrt{2\alpha'}}
\varepsilon^{\alpha\beta} \lambda^A_\alpha(1) \Gamma^{ijk}_{AB}  \lambda^B_\beta(2)C_{ijk}(3) \ee
which is non-zero even if $\vec{d}_1=\vec{d}_2=\vec{d}$ and $\vec{d}_3=-2\vec{d}$.

Notice that the doubly-charged state described by $\mathcal{V}_C$ may be taken to be near-BPS and as such long-lived if $|\vec{d}|>> \sqrt{\alpha'}$ and $g_s<<1$. Indeed in this case it may be considered as a low-lying string excitation on the BPS state represented by an open string with minimal length stretched between the two D3-branes (physical and image) associated to the $U(1)$ gauge group. This will play an important role in later considerations about the life-time of this and similar (higher-spin) states.

\subsection{ {4-pt amplitude}}

\begin{figure}[h!]
\centering
\includegraphics[scale=0.3]{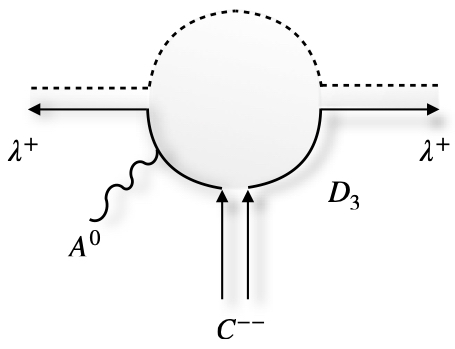}
\includegraphics[scale=0.3]{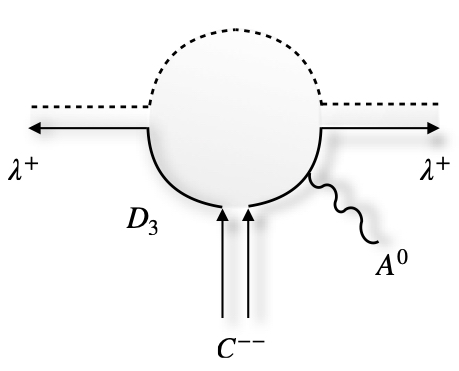}
\caption{Amplitude with one $U(1)$ photon, two massive BPS gaugini $\lambda$ and a massive non BPS scalar $C$.}
\label{Amp3+1}
\end{figure}

Let us then consider the 4-point amplitude with a single photon insertion, two massive singly charged BPS gaugini $\lambda^+$ and a massive non BPS doubly-charged scalar $C^{-2}$ $viz.$
\be
{\cal A}_{3{+}1}= {{g}}^{4}_{op}{\cal C}_{D2}\prod_{j=0}^{3} \int\frac{
dz_{j}}{V_{CKG}}\left<\mathcal{V}_{a}(k,z_{0})\mathcal{V}_{\lambda^{+}}(K_{1},z_{1})\mathcal{V}_{\lambda^{+}}(K_{2},z_{2})\mathcal{V}_{C^{-2}}(K_{3},z_{3})\right>
\label{amp1}
\ee
where ${\cal C}_{D_{2}}=(g_{op}2\alpha')^{-2}$. The two contributions ${\cal A}_{A\lambda^+\lambda^+C^{-2}}$ and ${\cal A}_{\lambda^+\lambda^+A C^{-2}}$  are shown in Fig.~\ref{Amp3+1}. 

The final result reads 
\be
{\cal A}_{3{+}1}={{g^{2}_{op}\over \sqrt{2\alpha'}}}
\lambda^A_\alpha(1) \Gamma^{ijk}_{AB}  \lambda^B_\beta(2)C_{ijk}(3)\left\{\left( \frac{a{\cdot} p_{1}\varepsilon^{\alpha\beta}{+}a{\cdot}\sigma^{\alpha\beta}{\cdot}k}{k{\cdot} p_{1}}-\frac{a{\cdot} p_{3}\varepsilon^{\alpha\beta}}{k{\cdot} p_{3}}\right)\times \right.\ee
$$\left.\frac{\Gamma(2\alpha' k{\cdot} p_{1}+1)\Gamma(2\alpha' k{\cdot} p_{3}+1)}{\Gamma(1-2\alpha' k{\cdot} p_{2})} + (1\leftrightarrow 2)\right\}$$
where $\sigma_{\mu\nu}^{\alpha\beta}$ is the Lorentz generator for chiral spinors. It is convenient to set $k=\omega n=\omega(1, \vec{n})$ and define the `scattering lengths' \cite{ABFMhet}
\be
{{\ell}}_{a}=2\alpha' np_{a}\qquad {\rm :} \qquad \sum_{a=1}^{3} {{\ell}}_{a} = 0
\label{elle} \: , 
\ee
since $\sum_{a=1}^{3} k{\cdot} p_{a} = -k^{2} = 0$, by momentum conservation. Then one can write 
\be
{\cal A}_{3{+}1} = {{g^{2}_{op}\over \sqrt{2\alpha'}}}\lambda^A_\alpha(1) \Gamma^{ijk}_{AB}  \lambda^B_\beta(2)C_{ijk}(3)\left\{
\left( \frac{a{\cdot} p_{1}\varepsilon^{\alpha\beta}{+}\omega a{\cdot}\sigma^{\alpha\beta}{\cdot}n}{{{\ell}}_{1}} {-} \frac{a{\cdot} p_{3}\varepsilon^{\alpha\beta}}{{{\ell}}_{3}}\right) {\cal H}_{13}(\omega) {+} (1\leftrightarrow 2) \right\}
\label{amp5}
\ee
where the analytic function
\be
\mathcal{H}_{13}(\omega) = {1\over \omega} \frac{\Gamma(1{+}\omega {{\ell}}_{1})\Gamma(1{+}\omega {{\ell}}_{3})}{\Gamma(1{-}\omega {{\ell}}_{2})}\label{funomega}
\ee
has simple poles at $\omega=0$ as well as at $\omega {{\ell}}_{1} = {-}n_{1}{-}1$ and  at $\omega {{\ell}}_{3} ={-}n_{3}{-}1$. The 
Mittag-Leffler (ML) expansion reads
\be
\label{MLexp}
\begin{aligned}
\mathcal{H}_{13}
=& \frac{1}{\omega} + \sum_{n_{1}=1}^{\infty}\frac{(-1)^{n_{1}}{{\ell}}_{1}}{n_{1}!( \omega {{\ell}}_{1} +n_{1})}\frac{\Gamma(1-n_{3}\lambda_{3,1})}{\Gamma(1+n_{1}\lambda_{2,1})} + \sum_{n_{3}=1}^{\infty}\frac{(-1)^{n_{3}}{{\ell}}_{3}}{n_{3}!( \omega {{\ell}}_{3} +n_{3})}\frac{\Gamma(1-n_{3}\lambda_{2,3})}{\Gamma(1+n_{3}\lambda_{1,3})}
\end{aligned}
\ee
where 
\be
\lambda_{{b},{a}}= \frac{{{\ell}}_{b}}{{{\ell}}_{a}} = {n{{\cdot}}p_b\over n{{\cdot}}p_a} = 
{k{{\cdot}}p_b\over k{{\cdot}}p_a}\,.
\ee
such that $\lambda_{3,1}{+}\lambda_{2,1}{=}{-}1$. One gets $\mathcal{H}_{23}(\omega)$  from $\mathcal{H}_{13}(\omega)$ after $1\leftrightarrow 2$ exchange. For the sub-leading spin term (due to the presence of the spin 1/2 gaugini) we also need the ML expansion of 
\be
\label{MLexp}
\begin{aligned}
\omega\mathcal{H}_{13} = \mathcal{H}^{\rm spin}_{13}
= &1{-}{\hspace{-1mm}}\sum_{n_{1}=1}^{\infty}\frac{(-1)^{n_{1}}}{(n_{1}{-}1)!( \omega {{\ell}}_{1}{ +}n_{1})}\frac{\Gamma(1{-}n_{3}\lambda_{3,1})}{\Gamma(1{+}n_{1}\lambda_{2,1})}{-}\sum_{n_{3}=1}^{\infty}\frac{(-1)^{n_{3}}}{(n_{3}{-}1)!( \omega {{\ell}}_{3} {+}n_{3})}\frac{\Gamma(1{-}n_{3}\lambda_{2,3})}{\Gamma(1{+}n_{3}\lambda_{1,3})}
\end{aligned}
\ee
that has no soft pole anyway. 

At large distances $R>>L$ one has  
\be 
\widetilde{{A}}^{\mu}(\omega, \vec{x})= g_{op}\,  
 {e^{{{i}} \omega R}\over 4\pi R} \sum_j {\,Q_j \over  n{\cdot}p_j} {\cal F}^\mu_j (\omega,\vec{n}; p_j) \,
\widehat{{\cal A}}_{3}(p_j) 
\ee
where 
\be 
\begin{aligned}
&{\cal F}^\mu_1= {\cal H}_{13}p_1^\mu\varepsilon_{\alpha\beta} {+}  {\cal H}^{spin}_{13} (n{\cdot}\sigma)^\mu_{\alpha\beta}, \,\,  {\cal F}^\mu_2= {\cal H}_{23}p_2^\mu\varepsilon_{\alpha\beta} {+}  {\cal H}^{spin}_{23} (n{\cdot}\sigma)^\mu_{\alpha\beta},\,\,   {\cal F}^\mu_3= {1\over 2}p_3^\mu\varepsilon_{\alpha\beta}({\cal H}_{13}+{\cal H}_{23})
\end{aligned}
\ee
Note that the `spin' terms  $J^{\mu\nu}_a= L^{\mu\nu}_a + S^{\mu\nu}_a$ act on the (non-zero) 3-point amplitude $\widehat{{\cal A}}^{\alpha\beta}_{3}$ of Eq. (\ref{nonzero3pt}) and some spin-average is understood in the unpolarised case. 
With this proviso we are ready to perform the FT in $\omega$ to the time coordinate {\it viz.}
\be 
\label{after}
{{A}}^{\mu}(t, \vec{x})=
{{g_{op}}\over 4\pi R} \sum_j {Q_j \over {n{\cdot}{p}_j}} \int_{-\infty}^{+\infty} {d\omega \over 2\pi } e^{- {{i}} \omega u} {\cal F}^\mu_j(\omega,{{\ell}}_j) \, .
\ee
The soft pole at $\omega{=}0$ {{EM}} is responsible for the EM memory (DC effect). The constant term in $\omega$ produces the spin correction to the EM memory effect \cite{NewMemo, PastPhysRep}, already discussed above, whereby the Lorentz generators act on the `hard' 3-point (non-zero unpolarised) amplitude. 
The infinite tower of poles on the real axis generate the superstring corrections $\Delta_s A^{\mu}(t, \vec{x})$.  

Deforming the integration path from the real line, {\it i.e.} $kp_a{\rightarrow} kp_a{-}{{i}} \epsilon$, as required by causality, and integrating in $\omega$ yield infinite series in the variables ${{\zeta}}_j = e^{iu/\ell_j}$, with $\ell_j=2\alpha' n p_j$, that expose the non-perturbative nature of the effect in $\alpha'$. As already mentioned before, in the usual low-energy expansion in powers of $\alpha'$ this effect would be completely hidden. 

Assuming the two BPS gaugini to be incoming $\lambda^+(p_1)$ and $\lambda^+(p_2)$ ($p^\mu{=}{-}p^\mu_{\rm phys}$)  and the photon $a(k)$ as well as the non-BPS scalar $C^{-2}(p_3)$ to be  outgoing ($p^\mu{=}{+} p^\mu_{\rm phys}$), one has ${{\ell}}_{1,2}>0$ and ${{\ell}}_3<0$. 

At $u=t-R >0$ one finds 
\be 
\begin{split}
\Delta^{(>)}_s{{A}}^{\mu}(t, \vec{x}) = -{{g} \over 4\pi R}&\Bigg\{ { 1\over n{{\cdot}}p_1} \sum_{n_3=1}^\infty 
{(-)^{n_3} \over n_3!} {\Gamma(1{-}n_3\lambda_{1,3})
\over  
\Gamma(1{+}n_3\lambda_{2,3})} (p_1^{\mu}  - {n_3\over \ell_3} nJ_1^\mu )e^{{{i}} n_3 u/{\ell}_3} \\
&+ {1 \over n{{\cdot}}p_2} \sum_{n_3=1}^\infty 
{(-)^{n_3} \over n_3!} {\Gamma(1{-}n_3\lambda_{2,3})
\over  
\Gamma(1{+}n_3\lambda_{1,3})} (p_2^{\mu}  - {n_3\over \ell_3} nJ_2^\mu )e^{{{i}} n_3 u/{\ell}_3}\\
&- { p_3^{\mu} \over n{{\cdot}}p_3} \sum_{n_3=1}^\infty 
{(-)^{n_3} \over n_3!} \left({\Gamma(1{-}n_3\lambda_{1,3})
\over  
\Gamma(1{+}n_3\lambda_{2,3})} {+}{\Gamma(1{-}n_3\lambda_{2,3})
\over  
\Gamma(1{+}n_3\lambda_{1,3})}  \right)\,e^{{{i}} n_3 u/{\ell}_3} \Bigg\}
\end{split}
\label{A>}
\ee
where $nJ^{\mu}_{j} = n_{\nu}J^{\nu\mu}_{j}$, while at $u=t-R <0$ one finds 
\begin{align}
\Delta^{(<)}_s{{A}}^{\mu}(t, \vec{x}) ={{g}\over 4\pi R} & \Bigg\{ { 1\over n{{\cdot}}p_1}
 \sum_{n_1=1}^\infty 
{(-)^{n_1} \over n_1!} {\Gamma(1{-}n_1\lambda_{3,1})
\over \Gamma(1{+}n_1\lambda_{2,1})} (p_1^{\mu}  - {n_1\over \ell_1} nJ_1^\mu )e^{{{i}} n_1 u/{{{\ell}}_1}}\nonumber \\ 
&+{1 \over n{{\cdot}}p_2}
 \sum_{n_2=1}^\infty 
{(-)^{n_2} \over n_2!} {\Gamma(1{-}n_2\lambda_{3,2})
\over \Gamma(1{+}n_2\lambda_{1,2})} (p_2^{\mu}  - {n_2\over \ell_2} nJ_2^\mu ) e^{{{i}} n_2 u/{{{\ell}}_2}}\label{A<} \\
&\hspace{-2cm}-{p_3^{\mu} \over n{{\cdot}}p_3} \left(
 \sum_{n_1=1}^\infty 
{(-)^{n_1} \over n_1!} {\Gamma(1{-}n_1\lambda_{3,1})
\over \Gamma(1{+}n_1\lambda_{2,1})} e^{{{i}} n_1 u/{{{\ell}}_1}} + \sum_{n_2=1}^\infty 
{(-)^{n_2} \over n_2!} {\Gamma(1{-}n_2\lambda_{3,2})
\over \Gamma(1{+}n_2\lambda_{1,2})} e^{{{i}} n_2 u/{{{\ell}}_2}}  \right)\Bigg\}\nonumber
\end{align}

The dependence on the position of the observer is coded in $\vec{n}{=}\vec{x}/R$, that appears in $n{{\cdot}}p_j{=}{-} E_j(1{-}\vec{n}\vec{v}_j)$ both in the denominators and in the exponents. Before trying and summing the above expressions, let us stress that in general they give contributions that are of the same order of magnitude as the `standard' QED memories, although they are modulated in $u$. Anyway they decay as $1/R$ and their `intensity' ${\cal I} \sim |A|^2\approx {\cal O}(1)$, in that there is no further $\alpha'$ suppression if $u\approx \ell_i$. In the following we will see how the non-vanishing spin of the colliding particles affects the string corrections to the EM memory effect.

\subsection{Closed-form expressions for Special Kinematics}
\label{SpecKin} 

The infinite series we found have finite radii of convergence and the physical values $|{{\zeta}}_{{j}}|{=}1$ may be outside the domain (circle) of analyticity.  However, the sum is possible in closed form for special kinematical regimes and analytic continuation can be explicitly carried out\footnote{The following kinematical considerations essentially reproduce the ones in similar contexts \cite{AAMBMFshort, ABFMhet}. We include them here for completeness. The knowledgeable reader may skip this section.}. 
 
Working in the CoM frame for clarity, one has $\vec{p}_1{=}\vec{p}{=}{-}\vec{p}_2$, $\vec{p}_3{=}{-}\vec{k}{=}{-}\omega\vec{n}$, so that 
\be
E_3{=} E^{\rm phys}_3{=}\sqrt{M_3^2{+}\omega^2}\quad , \quad E_{1,2}{=}{-}E^{\rm phys}_{1,2}{=}\sqrt{M_{1,2}^2{+}|\vec{p}|^2}\ee with  
\be
E^{\rm phys}_1 = {\widetilde{M}_3^2+M_1^2-M_2^2 \over 2\widetilde{M}_3} \,, \qquad E^{\rm phys}_2 = {\widetilde{M}_3^2+M_2^2-M_1^2 \over 2\widetilde{M}_3} \,, \qquad
|\vec{p}| = {\sqrt{{\cal F}(M^2_1,M^2_2,\widetilde{M}^2_3)}\over 2\widetilde{M}_3} \,,
\ee
where $\widetilde{M}_3 = E_3{+}\omega$ and ${\cal F}(x,y,z)= x^2+y^2+z^2-2xy-2yz-2zx$ is the ubiquitous `fake square'.  Setting 
\be
\mu_1={M_1^2\over \widetilde{M}_3^2}\quad , \quad \mu_2={M_2^2\over \widetilde{M}_3^2}\quad {\rm and} \quad \cos\theta =  {\vec{x}{\cdot} \vec{p}\over R|\vec{p}|}\ee
${\cal F}(M^2_1,M^2_2,\widetilde{M}^2_3)$ turns out to be positive when
$
0{<}\mu_1,\mu_2{<}1$ and $(\mu_1{-}\mu_2)^2{-}2(\mu_1{+}\mu_2){+}1{>}0$.

In the specific case under consideration, the two coalescing gaugini have the same mass thus $\mu_1=\mu_2=\mu=M^2/\widetilde{M}_3^2$ and 
\be
\lambda_{1,3} = -{1\over 2}+ {1\over 2} \cos\theta \sqrt{1- 4\mu} 
\, , \qquad \qquad  
\lambda_{2,3} = -{1\over 2} - {1\over 2} \cos\theta\sqrt{1- 4\mu} \,,
\ee
For instance, $\theta = \pi/2$ {\it i.e.} $\cos\theta = 0$ corresponds to $\ell_{1}=\ell_{2}=-\frac{1}{2}\ell_{3}$ so that  
\be
\lambda_{1,3}\equiv\lambda_{2,3}=-\frac{1}{2}\,,\quad\lambda_{3,1}\equiv\lambda_{3,2}=-2\,,\quad\lambda_{1,2}\equiv\lambda_{2,1}=1\,.
\label{lambdaSpec}
\ee
Using this particular choice of $\lambda_{a,b}$ in \eqref{A>} and \eqref{A<}, for $u>0$ one finds 
\be
\begin{aligned}
\Delta^{(>)}_s A^{\mu}(t, \vec{x}) &=\hspace{-1mm}{-}{{g} \over{4\pi}R}\left({p^{\mu}_1 {+} i\,nJ_1^\mu \partial_u\over n{\cdot}p_1} + {p^{\mu}_2 + i\,nJ_2^\mu \partial_u\over n{\cdot}p_2} - 2{p^{\mu}_3\over n{\cdot}p_3} \right)\sum_{n_3=1}^\infty 
{(-)^{n_3} \over n_3!} {\Gamma(1{+}{n_3\over2})
\over  
\Gamma(1{-}{n_3\over 2})} \,e^{{{i}} n_3 u/{\ell}_3}\\
&={{g} \over{4\pi}R}\left({p^{\mu}_1 + i\,nJ_1^\mu \partial_u\over n{\cdot}p_1} + {p^{\mu}_2 + i\,nJ_2^\mu \partial_u\over n{\cdot}p_2} - 2{p^{\mu}_3\over n{\cdot}p_3} \right){e^{{{i}} u/{\ell}_3} \over \sqrt{4 + e^{2{{i}}u/{\ell}_3}}}
\end{aligned} 
\ee
with $n{{\cdot}}p_j = - E_j\left(1-{\vec{x}\vec{v}_j\over{4\pi}R} \right)$, while for $u<0$ one finds 
\be
\begin{aligned}
\Delta^{(<)}_s A^{\mu}(t, \vec{x}) &={{g}\over{4\pi}R}\left({p^{\mu}_1 {+} i\,nJ_1^\mu \partial_u\over n{\cdot}p_1} + {p^{\mu}_2 + i\,nJ_2^\mu \partial_u\over n{\cdot}p_2} - 2{p^{\mu}_3\over n{\cdot}p_3} \right) \sum_{n_1=1}^\infty 
{(-)^{n_1} \over n_1!} {\Gamma(1{+}{2n_1})
\over \Gamma(1{+}n_1)} e^{{{i}} n_1 u/{{{\ell}}_1}}\\
&= {{g}\over{4\pi}R}\left({p^{\mu}_1 {+} i\,nJ_1^\mu \partial_u\over n{\cdot}p_1} + {p^{\mu}_2 + i\,nJ_2^\mu \partial_u\over n{\cdot}p_2} - 2{p^{\mu}_3\over n{\cdot}p_3} \right)\left({1 \over \sqrt{1+ 4e^{{{i}} u/{{{\ell}}_1}}}} - 1\right)
\end{aligned}
\ee
Notice that although the series have finite radii of convergence ($|{{\zeta}}_3|<2$, $|{{\zeta}}_1|<1/4$), the situation is similar to the open bosonic string. Once the series has been summed in closed form, the resulting expression is well define everywhere except for the branching points $\zeta_3= \pm 2 i$ and 
$\zeta_{1,2}= \pm i/4$. 
In fact planar duality entails $\Delta^{(>)}_s{{A}}^\mu+\Delta^{(<)}_s{{A}}^\mu=0$ very much as in \cite{AAMBMFshort}. The main difference with respect to the latter is the presence of extra terms due to the spin of the two gaugini. {As we will momentarily show, the effect of the spin may prove very significant on the amplitude and as a result on the string corrections to the EM memory.}

This affects both the `standard' EM memory, including the spin memory, and the string corrections thereof in the form of single derivatives wrt to $u \sim \log\zeta$. We will see later on that higher-spin states produce higher derivatives wrt to $u$. The main difference wrt to the EM memory is the modulation in $u$ with periods set by $\ell_j$.  
It is worth recalling that in the case of closed (heterotic) strings \cite{ABFMhet},  analytic continuation in the same kinematical regime as above produces log terms  that are not periodic in $u$.

\subsubsection{Results for different `rational'  kinematics and high energy regime}

The series encoding the string corrections to the EM memory can be summed for other `rational' values of the kinematical variables $\lambda_{a,b}$. 

\begin{figure}[h!]
\centering
\includegraphics[scale=0.4]{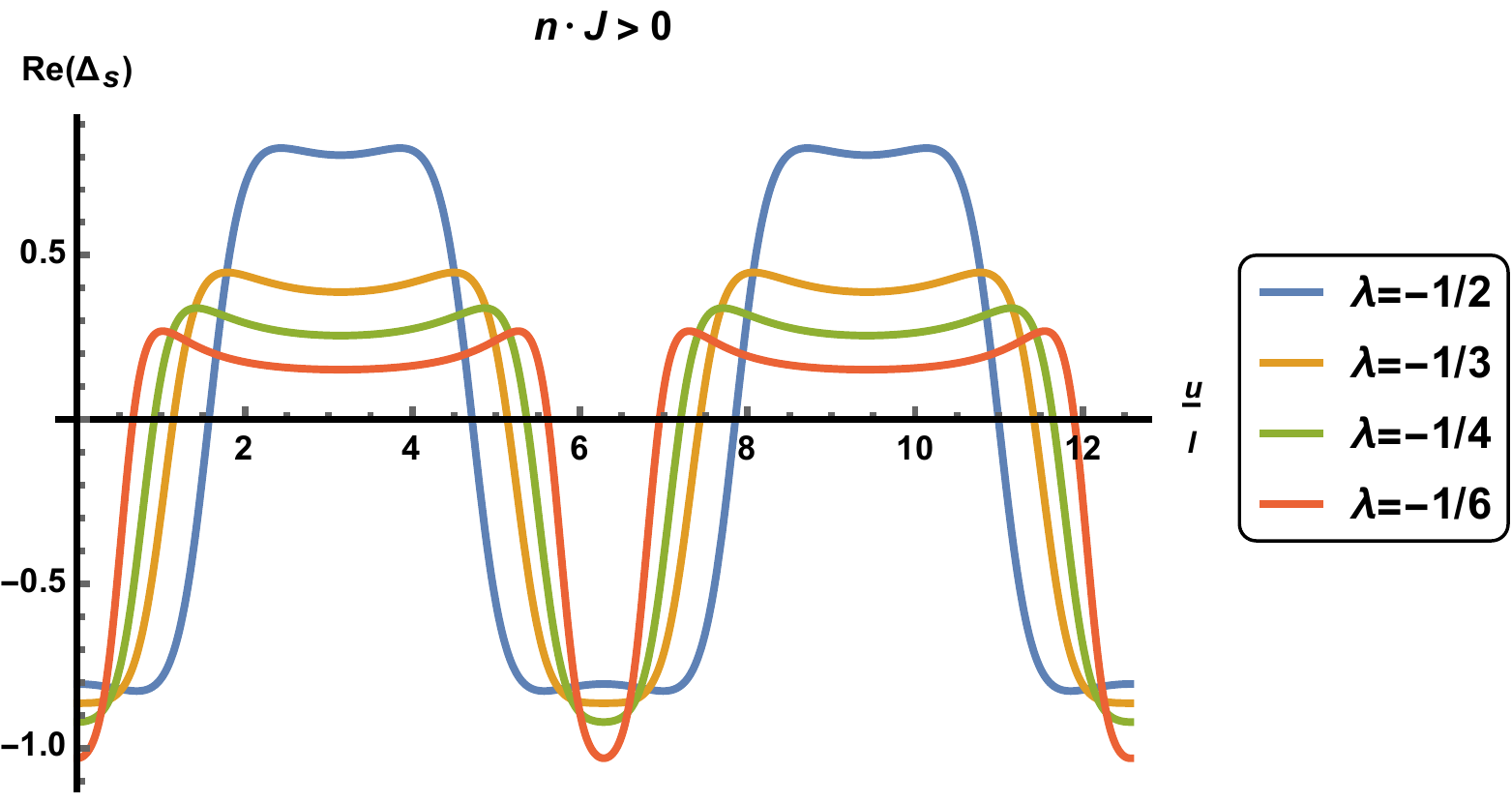}
\includegraphics[scale=0.4]{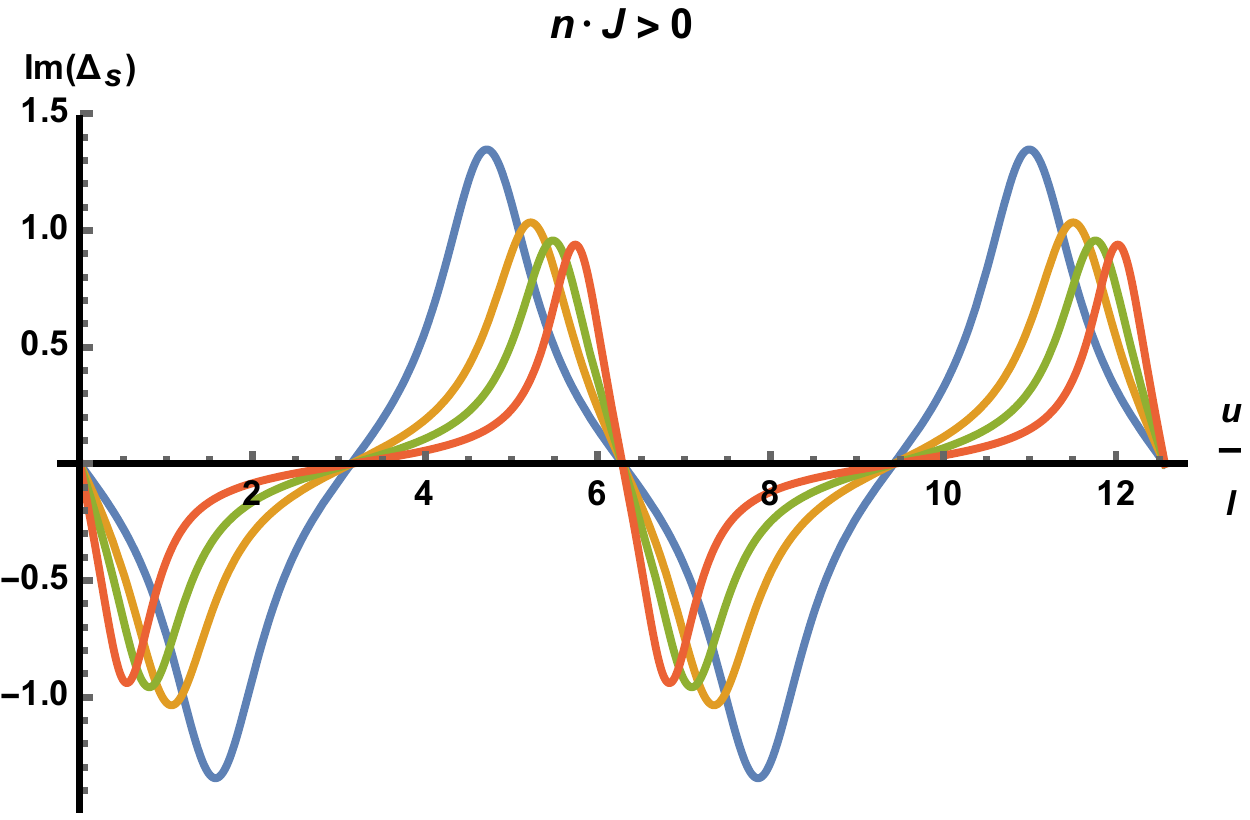}
\includegraphics[scale=0.4]{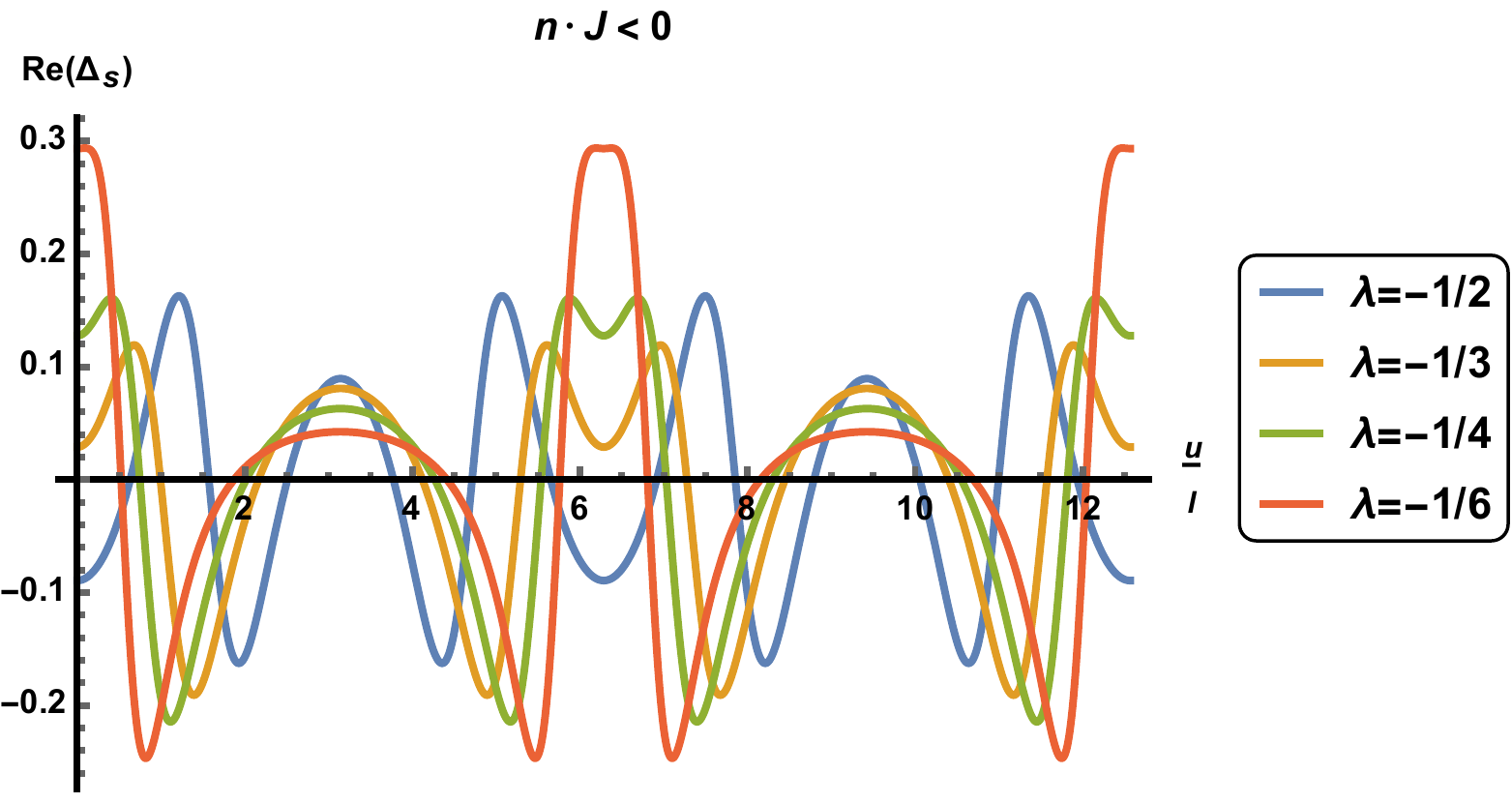}
\includegraphics[scale=0.4]{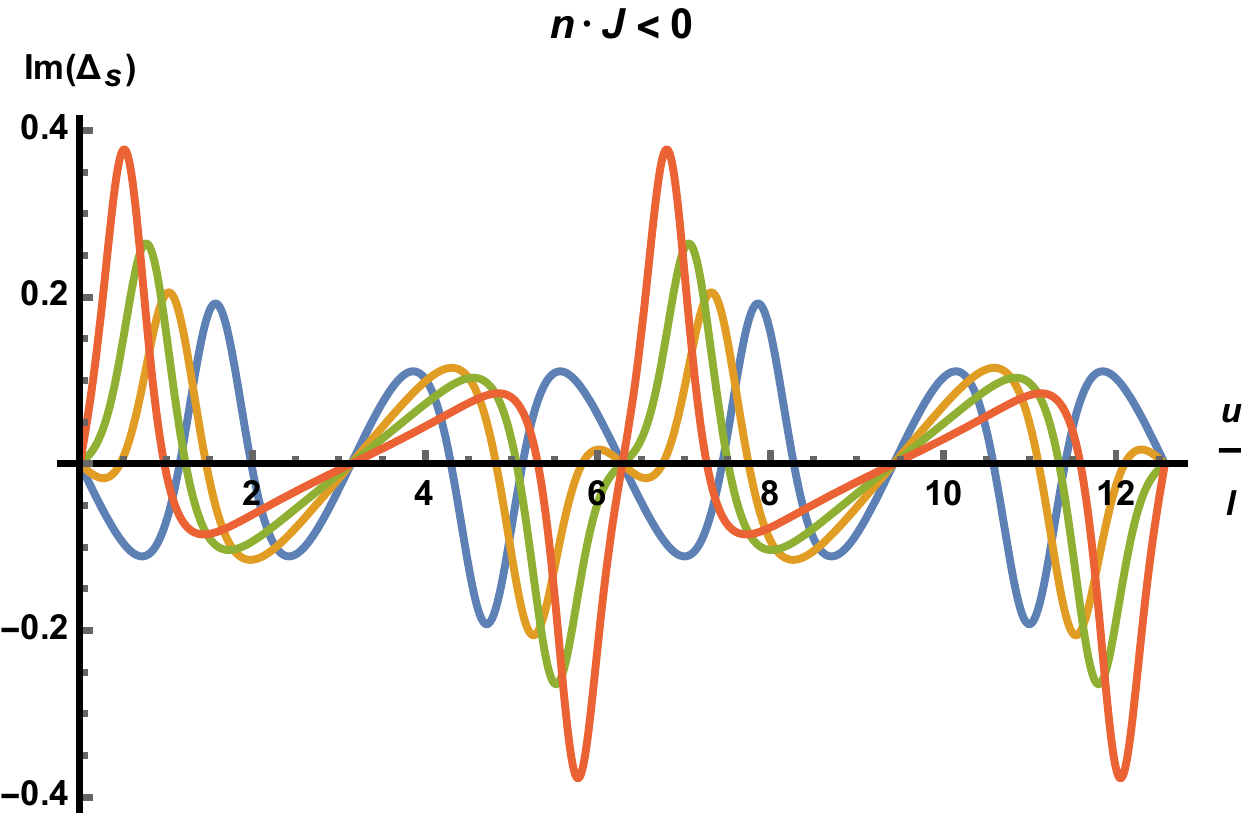}
\caption{Real and imaginary part of the Type I superstring correction to the EM memory for some rational kinematical values, with spin in two opposite directions ($nJ>0$ and $nJ<0$).}
\label{RatKinPlots1}
\end{figure}

In Table~\ref{RatKinTab}, we list a set of `rational' kinematical values used to plot the real and imaginary parts of the string corrections to the EM wave profile at fixed large $R$ as a function of $u/\ell$ in Fig.~\ref{RatKinPlots1}. {Notice the crucial role of the spin and in particular of the sign of $nJ$ on the EM wave-form. In principle the spin effect may be used to determine the spins of the colliding string states.} Clearly in order for the string effect to be measurable one needs {long-lived string states and} a time resolution $\Delta{t}\approx \ell\sim \alpha' E (1-\vec{n}{\cdot}\vec{v})$.
\begin{table}[h!]
\centering
\begin{tabular}{|c|c|c|c|c|c|}
\hline
 $\lambda_{13}$& $\lambda_{23}$ &$\lambda_{31}$&
 $\lambda_{21}$  & $\lambda_{12}$ &$\lambda_{32}$\\
\hline
-1/2 &-1/2& -2 &1& 1 &-2 \\
  -1/3 &-2/3& -3 &2& 1/2 &-3/2 \\
  -1/4 &-3/4& -4 &3& 1/3 &-4/3 \\
  -1/5 &-4/5& -5 &4& 1/4 &-5/4 \\
  -2/3 &-1/3& -3/2 &1/2& 2 &-3 \\
  -3/4 &-1/4& -4/3 &1/3& 3 &-4 \\
  \hline
\end{tabular}
\caption{\label{RatKinTab} Some examples of `rational' kinematical regimes.}
\end{table}

For instance taking the value $\lambda_{1,3}= -1/3$ and $\lambda_{2,3}= -2/3$, 
Eq.\eqref{A>} yields
\be
\begin{aligned}
&\Delta^{(>)}_s{{A}}^{\mu}(t, \vec{x}) 
={{g} \over 4\pi R}\Bigg\{\left[\frac{\left(p_{2}^{\mu} {+} i nJ_{2}^{\mu}\partial_{u}\right)}{n\cdot p_{2}} {-} \frac{p_{3}^{\mu}}{n\cdot p_{3}}\right]\Bigg[\frac{1}{3}\left(1{+}\frac{4}{27}e^{\frac{3iu}{\ell_{3}}}\right)^{-1/2}\\&\Bigg(\frac{1}{2^{1/3}}\left(1{+}\sqrt{1{+}\frac{4}{27}e^{\frac{3iu}{\ell_{3}}}}\right)^{1/3}{+}\frac{2^{1/3}}{3}e^{\frac{iu}{\ell_{3}}}\left(1{+}\sqrt{1{+}\frac{4}{27}e^{\frac{3iu}{\ell_{3}}}}\right)^{-1/3}\Bigg)\Bigg]
{+}\left[\frac{\left(p_{1}^{\mu} {+} i nJ_{1}^{\mu}\partial_{u}\right)}{n\cdot p_{1}} {-} \frac{p_{3}^{\mu}}{n\cdot p_{3}}\right]\\&\Bigg[\frac{2}{3}\left(1{-}\frac{4}{27}e^{\frac{3iu}{\ell_{3}}}\right)^{-1/2}\Bigg(\frac{2^{1/3}}{3}e^{\frac{iu}{\ell_{3}}}\left(1{+}\sqrt{1{-}\frac{4}{27}e^{\frac{3iu}{\ell_{3}}}}\right)^{-1/3}{-}\frac{1}{2^{1/3}}\left(1{+}\sqrt{1{-}\frac{4}{27}e^{\frac{3iu}{\ell_{3}}}}\right)^{1/3}\Bigg)\Bigg]\Bigg\}
\end{aligned}
\ee
while Eq.\,\eqref{A<} yields
\be
\begin{split}
&\Delta^{(<)}_s{{A}}^{\mu}(t, \vec{x}) ={{g} \over 4\pi R}\Bigg\{\left[\frac{\left(p_{1}^{\mu} + i nJ_{1}^{\mu}\partial_{u}\right)}{n\cdot p_{1}} - \frac{p_{3}^{\mu}}{n\cdot p_{3}}\right] \bigg[2\left({\sqrt{4+27
   e^{-\frac{3 i u}{\ell_{3}}}}}\right)^{-1}\\&\cosh \left(\frac{1}{3} \sinh ^{-1}\left(\frac{3\sqrt{3}
   }{2}e^{-\frac{3iu}{2\ell_{3}}}\right)\right) -1\bigg]{+}\left[\frac{\left(p_{2}^{\mu} + i nJ_{2}^{\mu}\partial_{u}\right)}{n\cdot p_{2}} {-} \frac{p_{3}^{\mu}}{n\cdot p_{3}}\right]\bigg[
  2\left({\sqrt{4-27
   e^{-\frac{3 i u}{\ell_{3}}}}}\right)^{-1} \\&\bigg[\cos \left(\frac{1}{6} \cos
   ^{-1}\left(1-\frac{27}{2} e^{-\frac{3 i u}{\ell_{3}}}\right)\right)-\sqrt{3} \sin
   \left(\frac{1}{3} \sin ^{-1}\left(\frac{3\sqrt{3}
   }{2}e^{-\frac{3iu}{2\ell_{3}}}\right)\right)\bigg]-1\bigg]\Bigg\}\,\,.
\end{split}
\ee
Otherwise starting for instance with the values $\lambda_{1,3}= -1/4$ and $\lambda_{2,3}= -3/4$, 
the result for Eq.\eqref{A>} is the following
\be
\begin{split}
&\Delta^{(>)}_s{{A}}^{\mu}(t, \vec{x}) ={{g} \over 4\pi R}\frac{e^{iu/\ell_{3}}}{4}\Bigg\{\left[3\frac{\left(p_{1}^{\mu}+inJ_{1}^{\mu}\partial_{u}\right)}{n\cdot p_{1}}+\frac{\left(p_{2}^{\mu}+inJ_{2}^{\mu}\partial_{u}\right)}{n\cdot p_{2}}-4\frac{p_{3}^{\mu}}{n\cdot p_{3}}\right]\\&\Bigg[\,_{3}F_{2}\left(\frac{1}{4},{7\over12},{11\over12};{1\over2},{3\over4};-\frac{27}{2^{8}}
   e^{\frac{4iu}{\ell_{3}}}\right) + \frac{5}{2^{5}}e^{\frac{2iu}{\ell_{3}}}\,_{3}F_{2}\left(\frac{3}{4},{13\over12},{17\over12};{5\over4},{3\over2};-\frac{27}{2^{8}}
   e^{\frac{4iu}{\ell_{3}}}\right)\Bigg]+\\&
-\frac{e^{\frac{iu}{\ell_{3}}}}{2}\left[3\frac{\left(p_{1}^{\mu}+inJ_{1}^{\mu}\partial_{u}\right)}{n\cdot p_{1}}{+}\frac{\left(p_{2}^{\mu}+inJ_{2}^{\mu}\partial_{u}\right)}{n\cdot p_{2}}{-}2\frac{p_{3}^{\mu}}{n\cdot p_{3}}\right]\Bigg[\,_{3}F_{2}\left(\frac{1}{2},{5\over6},{7\over6};{3\over4},{5\over4};-\frac{27}{2^{8}}
   e^{\frac{4iu}{\ell_{3}}}\right) \Bigg]\Bigg\}
\end{split}
\ee
while for Eq.\,\eqref{A<} the result reads
\be
\begin{split}
&\Delta^{(<)}_s{{A}}^{\mu}(t, \vec{x}) ={{g} \over 4\pi R}\bigg\{\left[\frac{\left(p_{1}^{\mu} + i nJ_{1}^{\mu}\partial_{u}\right)}{n\cdot p_{1}} - \frac{p_{3}^{\mu}}{n\cdot p_{3}}\right]\left[ \,
   _3F_2\left(\frac{1}{4},\frac{1}{2},\frac{3}{4};\frac{1}{3},\frac{2}{3};-\frac{2^{8}}{
   27}e^{-\frac{4iu}{\ell_{3}}
   }\right)-1\right]\\&+\frac{4}{3}\left[\frac{\left(p_{2}^{\mu} + i nJ_{1}^{\mu}\partial_{u}\right)}{n\cdot p_{2}} - \frac{p_{3}^{\mu}}{n\cdot p_{3}}\right] \bigg[\frac{5}{3}e^{-\frac{8iu}{3\ell_{3}}} \,
   _3F_2\left(\frac{11}{12},\frac{7}{6},\frac{17}{12};\frac{4}{3},\frac{5}{3};-\frac{2^{8}}{
   27}e^{-\frac{4iu}{\ell_{3}}}\right)\\&-e^{\frac{4iu}{3\ell_{3}}} \,
   _3F_2\left(\frac{7}{12},\frac{5}{6},\frac{13}{12};\frac{2}{3},\frac{4}{3};-\frac{2^{8}}{
   27}e^{-\frac{4iu}{\ell_{3}}}\right)-1\bigg]\bigg\}\,.
\end{split}
\ee

In addition to the special `rational' kinematical regimes, it is interesting to consider the ultra-high-energy limit $|u|<<|\ell_j|$. Two possible regimes can take place: fixed angle$|\ell_1|\approx |\ell_2|\approx |\ell_3|$ and Regge $|\ell_1|<< |\ell_2|\approx |\ell_3|$. We have already discussed these regimes in \cite{AAMBMFshort, ABFMhet}, to which we refer the interested reader for the detailed analysis. Nothing new and special happens in the Type I superstring case wrt the unoriented open bosonic string since the main character in the play is the scalar amplitude of the Veneziano model. {Yet the spin effect, that can be easily included at the relevant saddle-points for the amplitudes, may affect the very form of the EM wave profile quite significantly as shown above.}

One can generalise the analysis in various other directions. We will consider in turn the inclusion of spinning states, higher-points and higher-loops. More realistic (chiral) models with open and unoriented strings will be addressed later on, together with macroscopic semi-classical coherent states that may have the desired mass, spin and compactness for the effect to be detectable.

\section{ {EM memory with spinning string states}}
\label{Spin}  

In order to investigate the effect of higher spins in the simplest possible context, we move back to where we started from: unoriented open bosonic string. We focus on the $U(1){\times}O(1)$ model, built in \cite{AAMBMFshort}, that {\it mutatis mutandis} coincides with the Type I model in the previous section. We can consider an amplitude with one $U(1)$ photon, two singly-charged `tachyons' $\phi^+$ and a doubly-charged higher-spin state $H_N^{-2}$ in the first Regge trajectory. Later on we will sketch how the analysis needs to be modified for the superstring. 

The relevant open bosonic string vertex operators are 
\be
\begin{aligned}
&{\cal V}_{a}(k)= a{\cdot}i\partial X e^{ik{\cdot}X},\, \quad {\cal V}_{\phi}(K)=\sqrt{2\alpha'}e^{iK{\cdot}X}\\&{\cal V}_{H}=\sqrt{2\alpha'}^{1{-}N} i^N H_{\mu_1\mu_2...\mu_N}\partial X^{\mu_1}\partial X^{\mu_2}...\partial X^{\mu_N}e^{iK{\cdot}X}
\end{aligned}
\ee
Henceforth we set $H_N= \zeta^{\otimes N}$ for convenience. 
The disk amplitude  
\be
{\cal A}_{3{+}1}= {{g}}^{4}_{op}{\cal C}_{D2}\prod_{j=0}^{3} \int\frac{
dz_{j}}{V_{CKG}}\left<\mathcal{V}_{a}(k,z_{0})\mathcal{V}_{\phi^{+}}(K_{1},z_{1})\mathcal{V}_{\phi^{+}}(K_{2},z_{2})\mathcal{V}_{H^{-2}}(K_{3},z_{3})\right>
\label{amp1}
\ee
involves two possible colour-ordered contributions: ${\cal A}_{\phi\phi A H}$
and ${\cal A}_{A\phi\phi H}$. 
Focussing on the former, putting $2\alpha'=1$, and performing the Wick contractions yield
\be\label{eqhs}
{\cal A}_{3{+}1}= g_{op}^{4}C_{D_{2}}\,\int_{z_{2}}^{z_{3}}dz_{0}  \prod_{i<j\ne 1}z_{ij}^{p_{i}{\cdot}p_{j}}  \Bigg\{{a{\cdot}\zeta \over z^{2}_{03}} {N}
\Bigg(  \sum_{i\ne 1,3}{\zeta{\cdot}p_{i}\over z_{3i}} \Bigg)^{N{-}1}   {+} \sum_{j\ne 0,1}{a{\cdot}p_{j}\over z_{0j}} \Bigg(\sum_{i\ne 1,3}{\zeta{\cdot}p_{i}\over z_{3i}} \Bigg)^{N} \Bigg\}
 \ee
where  $p_{0}=k$ and $SL(2,R)$ invariance allows to put $z_{1}=\infty, z_{2}=1, z_{0}=z, z_{3}=0$. Integrating over $z$ one finds 
\begin{align}
&{\cal A}_{3{+}1}={g_{op}^{2}}\big(-\zeta{\cdot}p_{2}\big)^{N} \Bigg\{\left({a{\cdot}p_{2}\over k{\cdot}p_{2}}{-}{a{\cdot}p_{3}\over{k{\cdot}p_{3}}} \right) {\Gamma(1+ k{\cdot}p_{2}) \Gamma(1+ k{\cdot}p_{3}) \over \Gamma(1- k{\cdot}p_{1})}\,+ \\
&+\sum_{h=1}^{N} \begin{pmatrix} N \\ h \end{pmatrix}  \left({\zeta{\cdot}k \over \zeta{\cdot}p_{2}  } \right)^{h}\Bigg[ {a{\cdot}p_{2}\over k{\cdot}p_{2}} -  {a{\cdot}p_{3}\over k{\cdot}p_{3} {-}{h\over 2\alpha'}} +  {a{\cdot}\zeta \over k{\cdot}\zeta}  {h \over (k{\cdot}p_{3} {-}{h} )}  \Bigg] {\Gamma(1{+}k{\cdot}p_{2}) \Gamma(1{+} k{\cdot}p_{3}{-}h) \over \Gamma(1{-} k{\cdot}p_{1}{-}h)} \Bigg\}\,.\nonumber
\end{align}
As for the spin 1/2 fermions in the Type I model discussed previously, the effect of the spin of $H_N$ is to produce derivatives wrt to the retarded time $u=t{-}R$. In order to show this, let us expand the amplitude \`a la Mittag-Leffler. There are various series of poles in $\omega$.

Focusing on the first term proportional to $a{\cdot}p_{2}$, the poles at $\omega \ell_{3}+1-h=-n_{3}$ yield an expression of the form
\be
{\cal C}_{a{\cdot}p_{2}}=\sum_{h=1}^{N}\begin{pmatrix} N \\ h \end{pmatrix}  
\left({ n{\cdot}\zeta \over \ell_{3}\, \zeta{\cdot}p_{2} } \right)^{h}  {d^{h{-}1}\over dx^{h{-}1}}   \sum_{n_{3}=0}^{\infty}{(-)^{n_{3}}\over n_{3}!} {\Gamma\big(1{-}\lambda_{23}(n_{3}{+}1{-}h)\big)\over \Gamma\big(1{+}\lambda_{13}(n_{3}{+}1{-}h){-}h\big) } e^{x(n_{3}{+}1{-}h)}
\ee
where $x=i u /\ell_{3}$, that looks rather intricate. Yet for special kinematics one can write the result in a compact form. For instance  for $\lambda_{23}=\lambda_{13}=-1/2$ one finds
\be
{\cal C}_{a{\cdot}p_{2}} = \sum_{h=1}^{N}\begin{pmatrix} N \\ h \end{pmatrix}\left(n{\cdot}\zeta \over \ell_{3}\, \zeta{\cdot}p_{2}\right)^{h}  {d^{h{-}1}\over dx^{h{-}1}} \left(  {e^{(1{-}h)x}\over 2^{h}(4{+}e^{2x})^{3/2}}  {\big[ 4(h{-}1) + h e^{x}\big( \sqrt{4{+}e^{2x}} {-} e^{x}\big) \big] \over \big(e^{x} {+}\sqrt{4{+}e^{2x}}\big)^{-h} }\right)
\ee
 Note that the presence of $\ell_{3}=\sqrt{2\alpha'}n{\cdot}p_{3}$  in the denominator should not be worrisome. String corrections to the memory effects are visible in the high-energy limit  $\sqrt{2\alpha'}n{\cdot}p_{3}\approx 1$. In the `usual' low-energy regime the effect is hidden, since all the poles and the resulting corrections would be absent.
\,The analysis of the other terms follows along the same way, for the pole at $\omega \ell_{2}+1=-n_{2}$ as well as for the pole at $1+\omega \ell_{1}=-n_{1}$.

\begin{figure}[h!]
\centering
\includegraphics[scale=0.5]{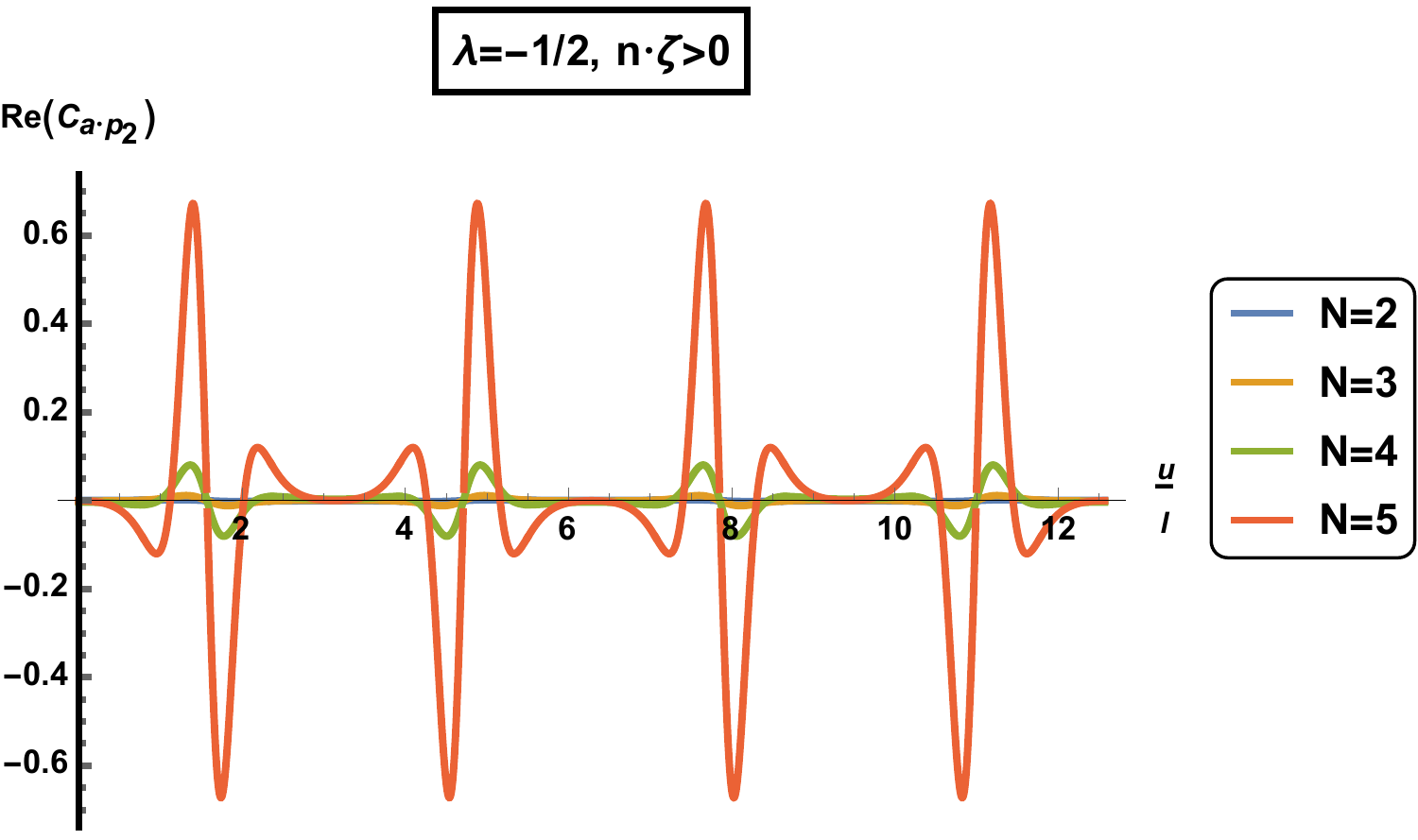}
\includegraphics[scale=0.5]{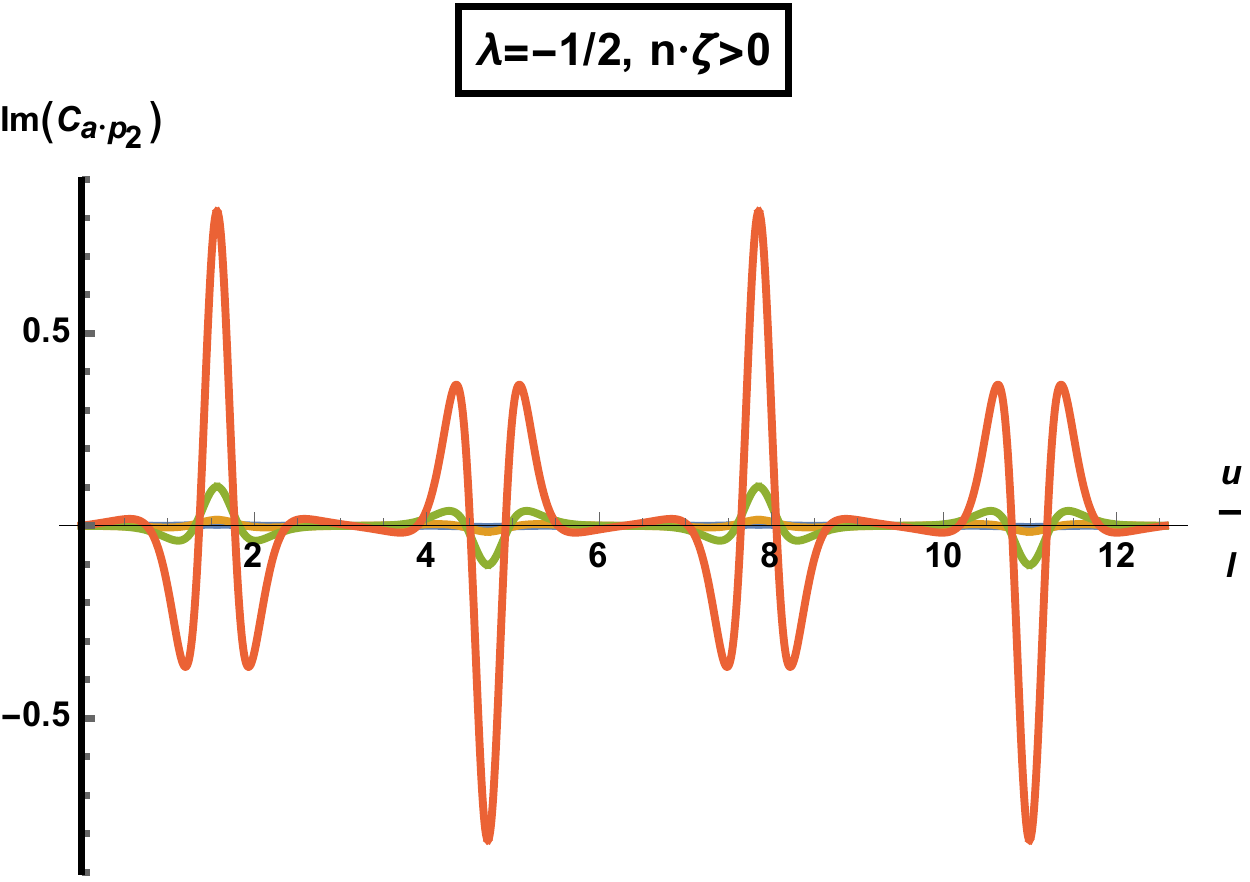}
\includegraphics[scale=0.5]{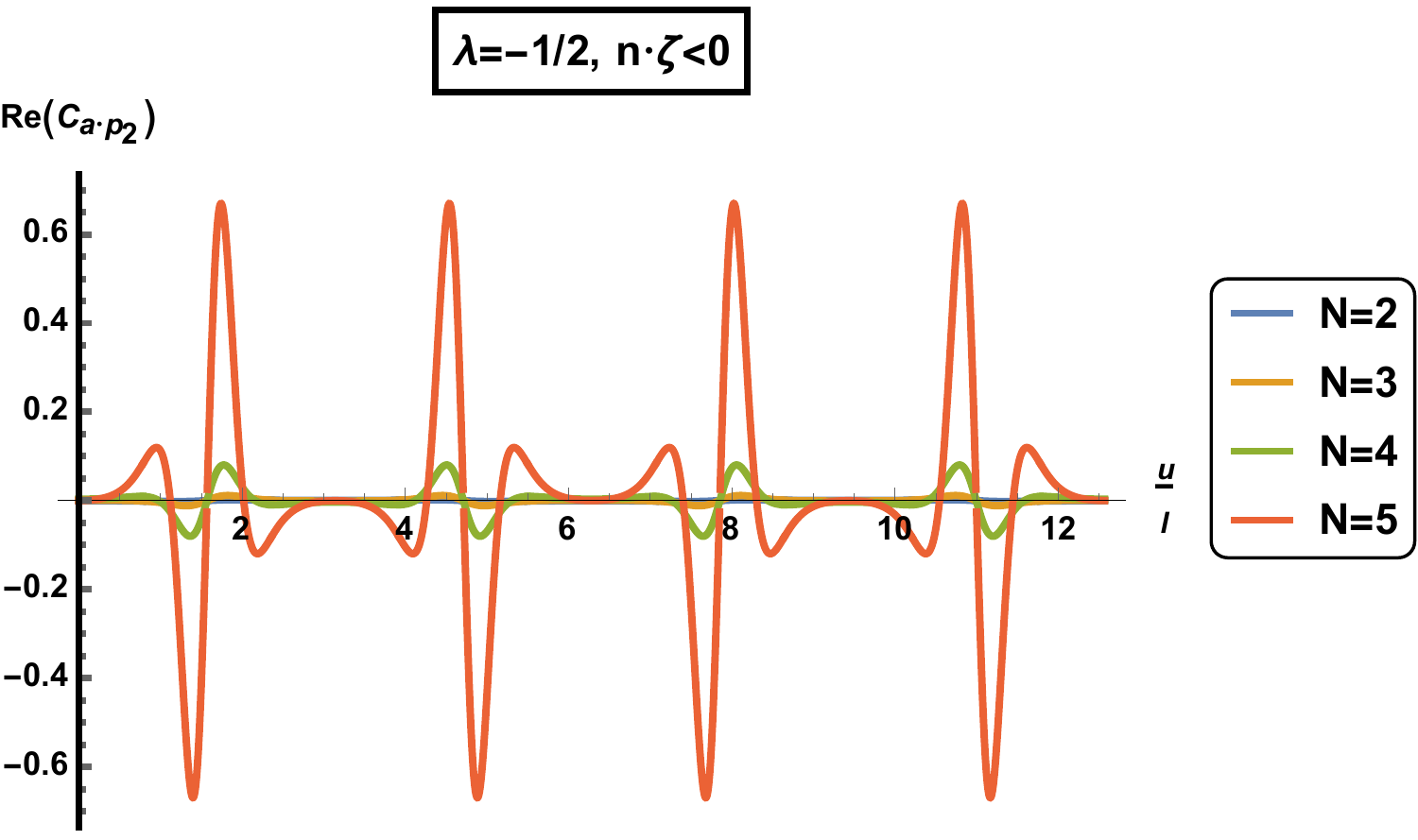}
\includegraphics[scale=0.5]{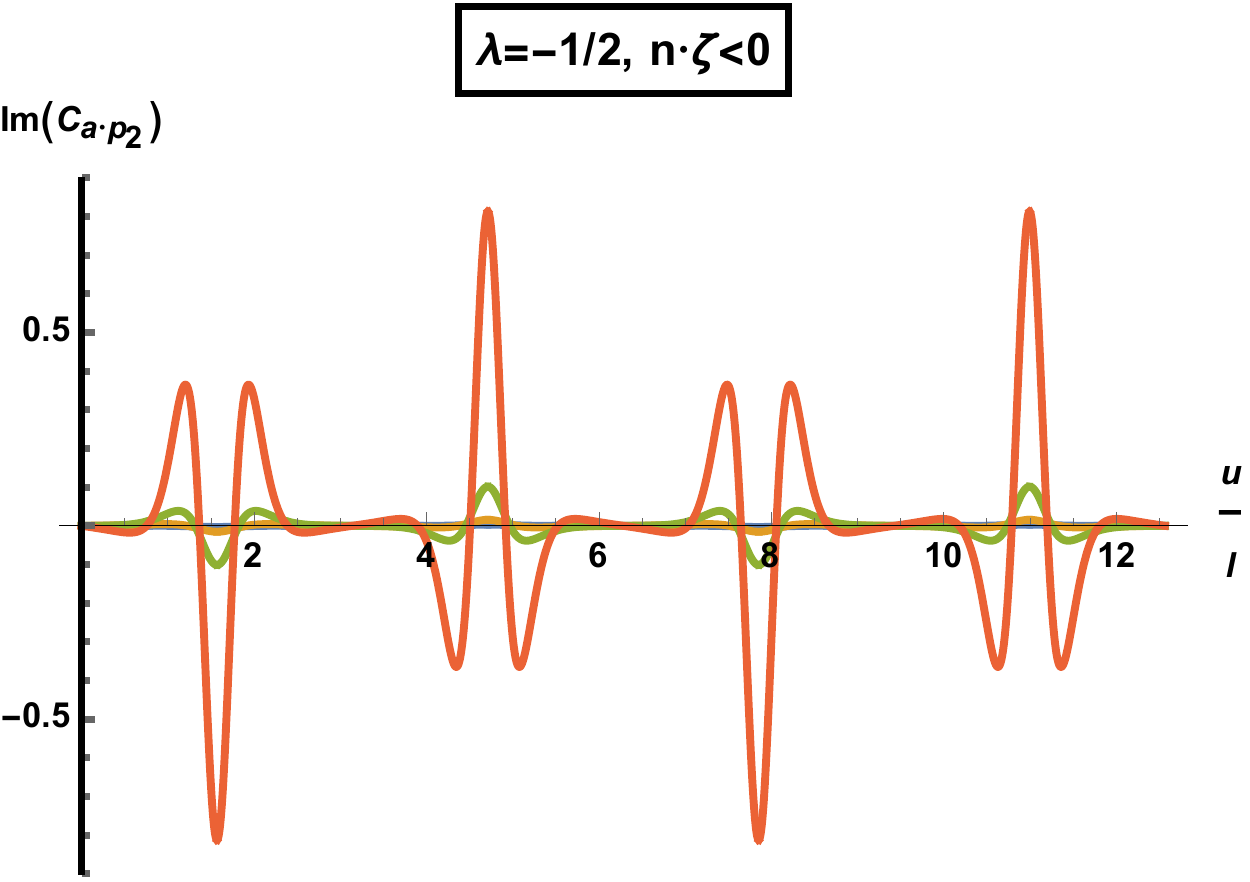}
\includegraphics[scale=0.5]{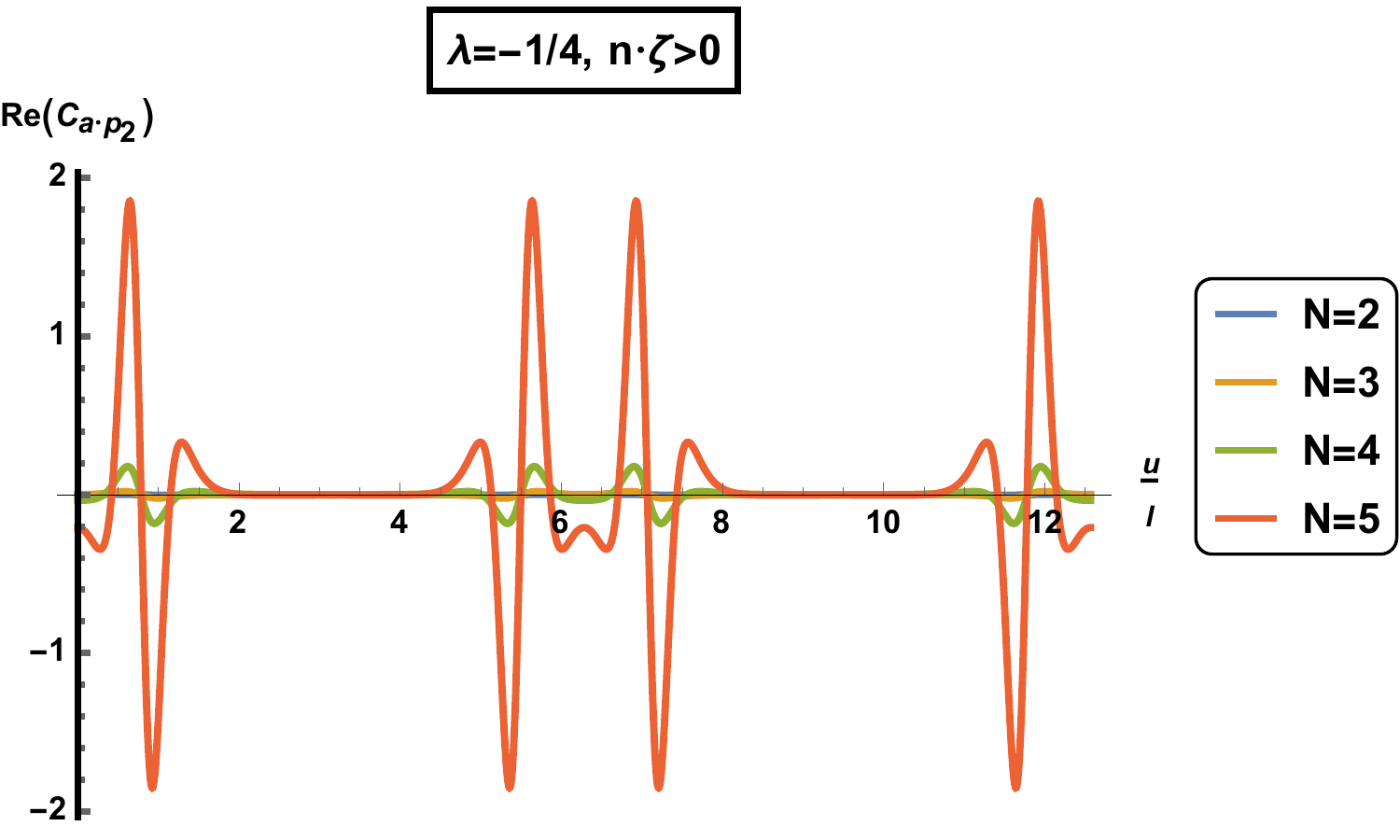}
\includegraphics[scale=0.5]{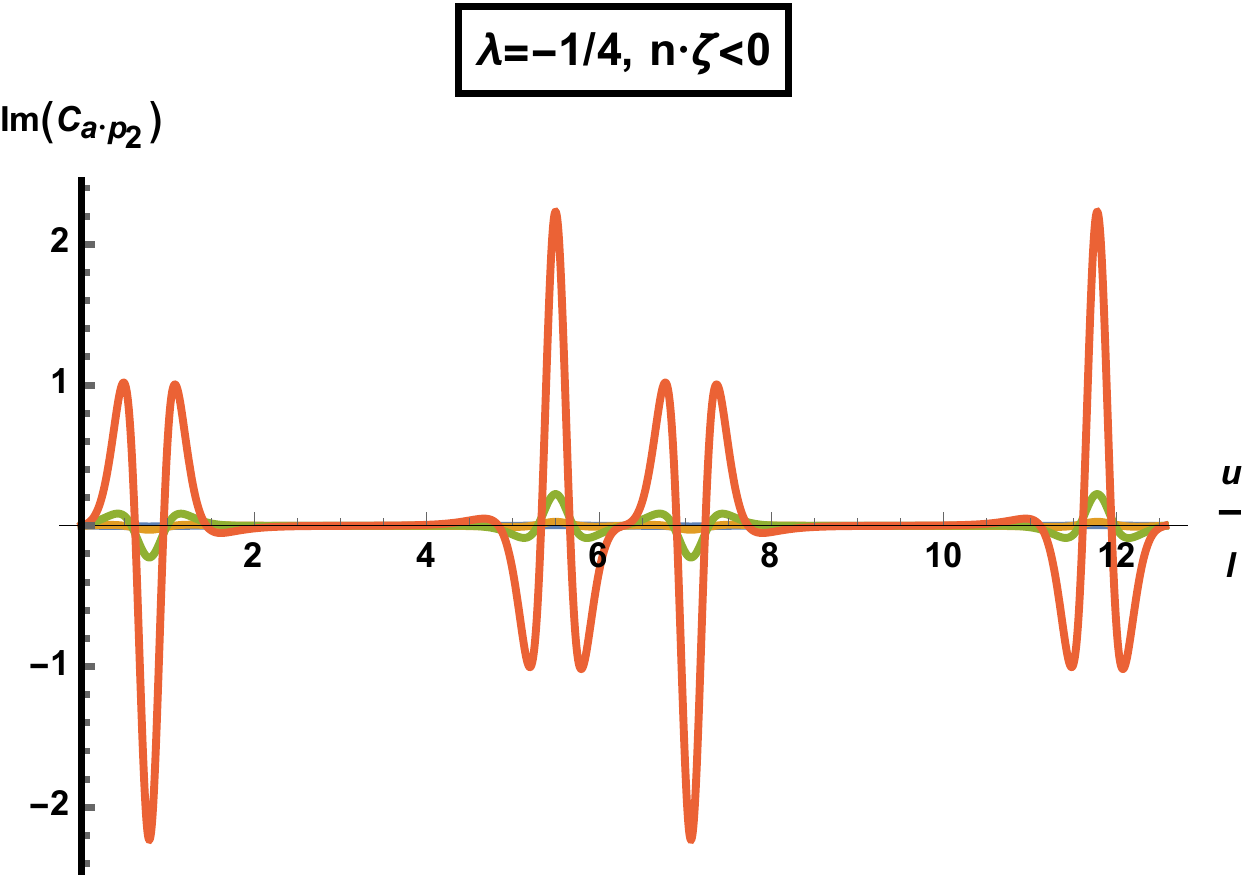}
\includegraphics[scale=0.5]{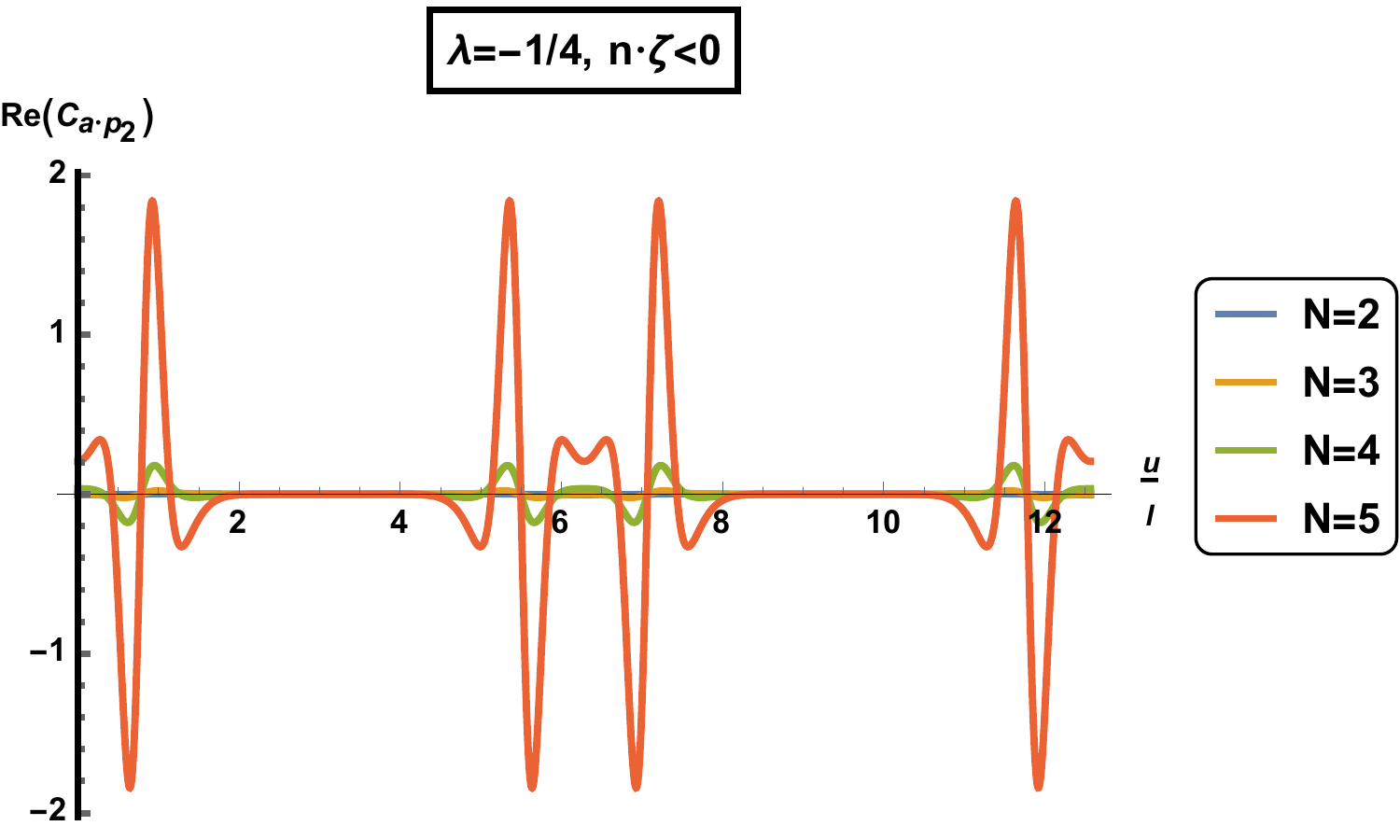}
\includegraphics[scale=0.5]{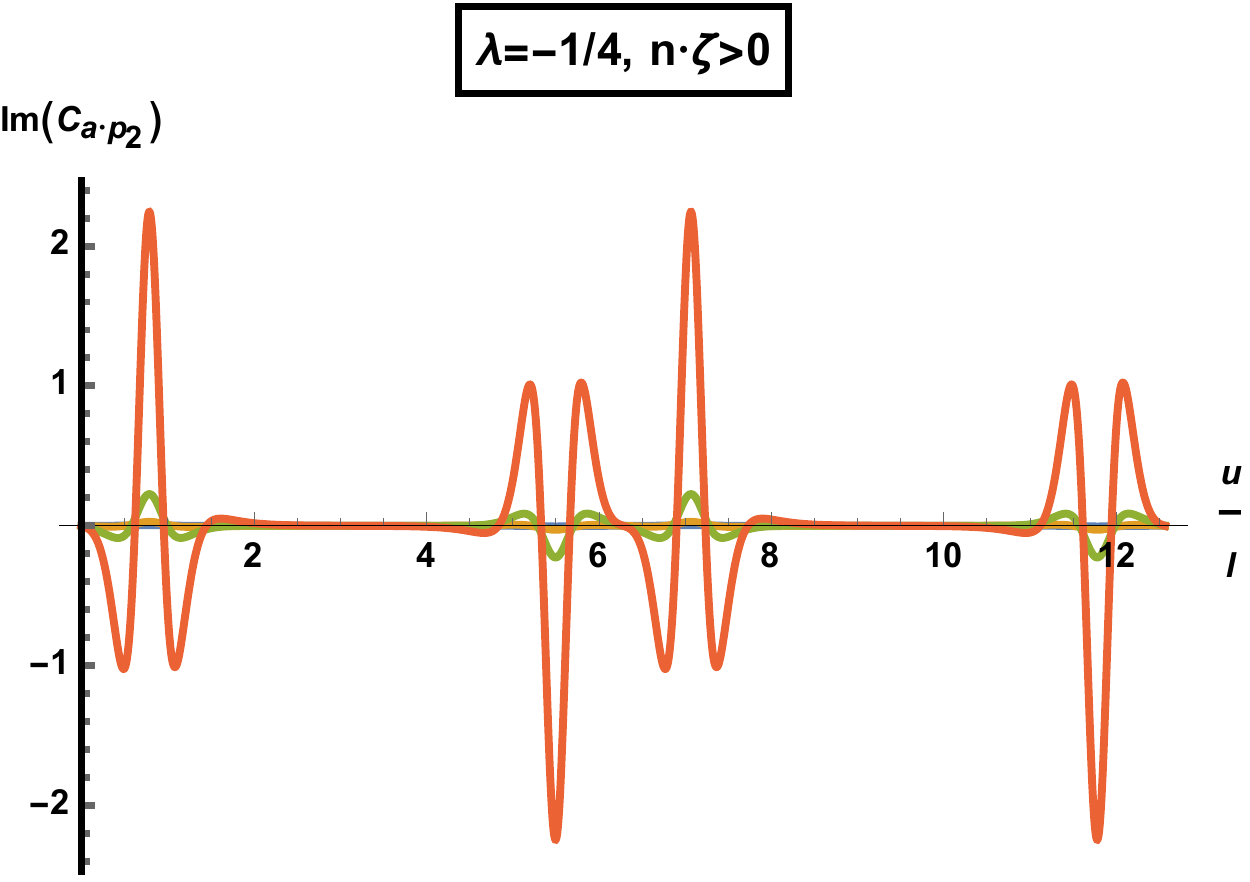}
\caption{Real and imaginary part of the string corrections to the EM spin memory for $\lambda=-1/2, \,-1/4$ rational kinematical values, with two opposite directions of $n{\cdot}\zeta \simeq 10^{-3}$. For each choice of $\lambda$ there are four corresponding profiles that differ for the spin values $N=2,3,4,5$, respectively.}
\label{RatKinPlots1Spin}
\end{figure}

The structure of the other contributions to the amplitude is similar. The lowest terms with $h =1$ reproduce the straight-forward generalisation of the spin memory, upon anti-symmetrization of $a$ and $n= k/\omega$. Higher derivative terms in $u$ up to $N^{\rm th}$ order are peculiar to string theory and to the specific higher-spin state considered (first Regge trajectory). The overall effect is quite dramatic as visible in Fig.\ref{RatKinPlots1Spin}, where the EM wave profiles are plotted for different values of $N$ and $n{\cdot}\zeta$.

One simple extension to the above analysis is to consider higher-spin states in the first Regge trajectory for Type I superstring compactified on tori \cite{Bianchi:2010es}. In the N-S sector the relevant vertex operator is
\be
V_{H_S}^{NS} = {\sqrt{2\alpha'}^{2-N}}H_{\mu_1\mu_2...\mu_N} e^{-\varphi} \psi^{\mu_1} \partial X^{\mu_2} ... \partial X^{\mu_N} e^{iPX}
\ee
with $P^\mu H_{\mu_1\mu_2...\mu_N}=0$ and $\eta^{\mu\nu} H_{\mu_1\mu_2...\mu_N}=0$.
In the R sector the relevant vertex operator is
\be
V_{\Lambda_S}^{R} =  {\sqrt{2\alpha'}^{1-N}}\Lambda^{\alpha A}_{\mu_1\mu_2...\mu_N} e^{-\varphi/2} S_\alpha \Sigma_A  \partial X^{\mu_1} ... \partial X^{\mu_N} e^{iPX}
\ee
with $(P^\mu \sigma^\mu_{\alpha\dot\alpha} + ...) \Lambda^{\alpha A}_{\mu_1\mu_2...\mu_N}=0$ and $\eta^{\mu\nu} \Lambda_{\mu_1\mu_2...\mu_N}=0$.

The computation of the amplitude with a single photon insertion, two massive BPS gaugini and a bosonic higher-spin state (with an internal $\psi^i$ replacing $\psi^{\mu_1}$, for simplicity) is straightforward, though tedious. The result is largely fixed by Lorentz invariance and R-symmetry. As in the bosonic string case, it
involves two possible colour-ordered contributions: ${\cal A}_{\lambda\lambda A H}$
and ${\cal A}_{A\lambda\lambda H}$. 
Focussing on the former, setting $H'_N= \chi_i{\otimes}\zeta^{{\otimes}N}$, putting $2\alpha'=1$ and performing the Wick contractions yields
\be
\begin{aligned}
&{\cal A}_{\lambda\lambda A H}=g_{op}^{4}C_{D_{2}} \chi_{AB}(3) \lambda^{\alpha A}(1) \lambda^{\beta B}(2)\int_{z_{2}}^{z_{3}}dz_{0}  \prod_{i<j\ne1}z_{ij}^{p_{i}{\cdot}p_{j}} \\
&\left\{ \varepsilon_{\alpha\beta} \left[
{a{\cdot}\zeta \over z^{2}_{03}} {N}
\Bigg( \sum_{i\ne 1,3}{\zeta{\cdot}p_{i}\over z_{3i}} \Bigg)^{N{-}1}{+} \sum_{j\ne 1,0}{a{\cdot}p_{j}\over z_{0j}} \left(\sum_{i\ne 1,3}{\zeta{\cdot}p_{i}\over z_{3i}} \right)^{N} \right]+ {a{\cdot} \sigma_{\alpha\beta}{\cdot} k \over z_{02}} \left( \sum_{i\ne 1,3}{\zeta{\cdot}p_{i}\over z_{3i}} \right)^{N} \right\}
\end{aligned}
\ee
where $p_{0}=k$ and $SL(2,R)$ invariance allows to put $z_{1}=\infty, z_{2}=1, z_{0}=z, z_{3}=0$. Integrating over $z_0$ produces an amplitude which is almost identical to the one in Eq.\,\eqref{eqhs} for the bosonic string, were it not for the absence of the tachyon pole and for the presence of the spin operators acting on the gaugini. {Clearly for $N>>1/2$ the effect of the latter will prove subdominant. Yet, in order for the  the present analysis to be reliable the superstring state should be near-BPS and long-lived. This requires $|\vec{d}|>> \sqrt{\alpha'}$ and $N$ not too large, {\it viz.} $N<< |\vec{d}|/\sqrt{\alpha'}$.} We refrain to display the final expression that is rather cumbersome and does not add much to what we already learned from the bosonic string. This  confirms that the higher-spin superstring corrections to the EM memory enjoy the same structure, {\it i.e.} they produce higher derivatives in $u$ acting on the original scalar amplitude ($N=0$) up to order $N$, the spin of the higher-spin state involved in the process. 

\section{Amplitudes at {n-points} and higher-loops}
\label{Npoint} 

Higher-points amplitudes show new kind of string corrections that we have already discussed in \cite{AAMBMFshort} for the bosonic string. Here we consider the same issue in the Type I context and highlight the main differences\footnote{The reader familiar with the bosonic case may only give a cursory look at this section.}. $U(1)$ charge conservation allows for amplitudes  
\be
{\cal A}[A_1, ..., A_{n_A}; \lambda^+_1,...,\lambda^+_{n_1}; \lambda^-_1,...,\lambda^-_{\bar{n}_1}; \chi^{+2}_1,...,\chi^{+2}_{n_2}; \chi^{{-}2}_1,...,\chi^{{-}2}_{\bar{n}_2}]
\ee
with $n_{1}{-}\bar{n}_{1}{+}2n_{2}{-}2\bar{n}_{2}{=}0$. Due to `twist' symmetry, $n_A$ must be even if $n_{1}{=}\bar{n}_{1}{=}n_{2}{=}\bar{n}_{2}{=}0$. Further constraints arise from supersymmetry. 


For amplitudes with a single photon, a soft pole $1/2\alpha' kp_{{{j}}}$ and 
a series of `massive' poles $1/2\alpha' kp_j{+}n$ arise from the region of integration where  the photon vertex is close to an adjacent (charged) vertex. At higher points there are 
multi-particle channels that arise from the region of integration where the photon is inserted on an internal leg. In this case there are no soft poles for generic hard momenta, since
\be
{(k+p_1+ .... p_m)^2 + M^2} = {2k(p_1+ .... p_m) +(p_1+ .... p_m)^2 + M^2}
\ee
Additional massive poles for $2\omega n P = {-}P^2{-}M^2$ with $P=\sum_{i=1}^m p_i$ appear that look sub-leading in (soft) $\omega$
\be
{1\over 2k(p_1+ .... p_m) +(p_1+ .... p_m)^2 + M^2} \approx {1\over (p_1+ .... p_m)^2 + M^2} - {2\omega n(p_1+ .... p_m)\over [(p_1+ .... p_m)^2 + M^2]^2} + ....
\ee
Integration over $\omega$ yields a new series of terms in ${{\zeta}}_P =\exp{i u\over 2\alpha' nP}$, from the tower of intermediate states, which represents a new feature w.r.t. to the 4-point case.

\begin{figure}
\centering
\includegraphics[scale=0.35]{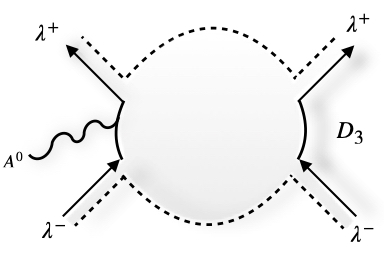}
\includegraphics[scale=0.35]{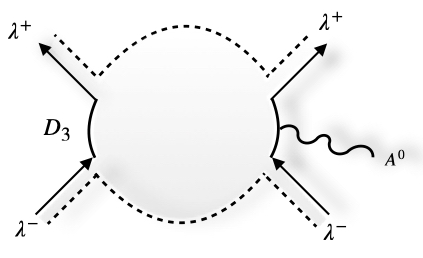}
\caption{Amplitude with one $U(1)$ photon, two massive BPS gaugini $\lambda$ and a massive non BPS scalar $C$.}
\label{amp5pt}
\end{figure}

Let us consider for instance the 5-point amplitude in Fig.~\ref{amp5pt}
\be
{\cal A}_{4{+}1}[\lambda^+(1), A(k), \bar\lambda^-(2), \bar\lambda^+(3), \lambda^-(4)] 
\ee
where the BPS anti-gaugino vertex operator\footnote{Since ${\cal N}=4$ multiplets (both massless and massive) are non-chiral, gaugini with same charge appear with both chiralities. With the chosen charge assignment $K_{\lambda^+} = (p^\mu, \vec{d})$ while $K_{\bar\lambda^-} = (p^\mu,-\vec{d})$.}  reads
\be
V_{\bar\lambda^{\pm}} = \sqrt{2\alpha'}\bar\lambda_A^{\dot\alpha} C_{\dot\alpha} \bar\Sigma^A e^{-\varphi/2} e^{iKX}
\ee
with $C_{\dot\alpha}$ and $\bar\Sigma^A$ spin fields with opposite chirality wrt $S^{\alpha}$ and $\Sigma_A$. The relevant correlators are
\be
\langle S^\alpha \Sigma_A e^{-\varphi/2}(4) C_{\dot\alpha} \bar\Sigma^C e^{-\varphi/2}(3)
S^\beta \Sigma_B e^{-\varphi/2}(2) C_{\dot\beta} \bar\Sigma^D e^{-\varphi/2}(1) \rangle = 
{\varepsilon^{\alpha\beta} \bar\varepsilon_{\dot\alpha\dot\beta}\over z_{13}z_{24}} 
\left({\delta_A^C \delta_B^D\over z_{12}z_{34}} + {\delta_A^D \delta_B^C\over z_{14}z_{23}}\right)
\label{interes}
\ee
and
\be
\langle \psi^\mu\psi^\nu(0) S_\alpha \Sigma_A e^{-\varphi/2}(4) C_{\dot\alpha} \bar\Sigma^C e^{-\varphi/2}(3)
S_\beta \Sigma_B e^{-\varphi/2}(2) C_{\dot\beta} \bar\Sigma^D e^{-\varphi/2}(1) \rangle =\ee
$$
\left({\sigma^{\mu\nu}_{\alpha\beta} \bar\varepsilon_{\dot\alpha\dot\beta} \over z_{04}z_{02} z_{13}} 
+ {\varepsilon_{\alpha\beta} \bar\sigma^{\mu\nu}_{\dot\alpha\dot\beta} \over z_{01}z_{03} z_{24}}\right)
\left({\delta_A^C \delta_B^D\over z_{12}z_{34}} + {\delta_A^D \delta_B^C\over z_{14}z_{23}}\right)
$$

Setting $z_4=\infty$, $z_3=0$ and 
$z_1=1$ with $0<z_A= z<1$ and $0<z_2= yz<z$ ({\it i.e.} $0<y<1$) one finds
\be
{\cal A}^{\rm c.o.}_{4{+}1}[1^+, a(k), 2^-,3^+,4^-]= g_{op}^3 \int_0^1 dz \int_0^1 dy \left({\delta_A^C \delta_B^D\over 1-yz} + {\delta_A^D \delta_B^C\over yz}\right) \ee
$$
\left[ \varepsilon^{\alpha\beta} \bar\varepsilon_{\dot\alpha\dot\beta}
\left( {a{{\cdot}}p_1\over 1-z}  - {a{{\cdot}}p_2\over z(1-y)} -{a{{\cdot}}p_3\over z}\right)
+ \left({\sigma^{\mu\nu}_{\alpha\beta} \bar\varepsilon_{\dot\alpha\dot\beta} \over z(1-y)}
- {\varepsilon_{\alpha\beta} \bar\sigma^{\mu\nu}_{\dot\alpha\dot\beta} \over z(1-z)}\right)\right] 
{\times}
$$
$$
(1-z)^{kp_1} (1-yz)^{ P_1P_2} (1-y)^{kp_2} 
z^{ (kp_2+kp_3+P_2P_3)+1} y^{P_2P_3} \, .
$$
Notice the appearance of the spin operators acting on the spin 1/2 fermions.
Using \be
(1-yz)^{P_1P_2-r} = \sum_{N=0}^\infty {(r-P_1P_2)_N\over N!}  y^N z^N 
\label{expan1}
\ee
with $r=0,1$ allows to integrate over $z$  and $y$ independently and get combinations of the form
\be 
{\cal F}_N(k,p_j) = {\Gamma(P_2P_3 {+} k(p_{2}{+}p_{3}){+}N{+}1)\Gamma(kp_1{+}1)
\over \Gamma(P_1P_4{-}kp_{3}{+}N{+}1)} {\Gamma(P_1P_4{+}k(p_{1}{+}p_{4}){+}N{+}1)\Gamma(kp_2{+}1)
\over \Gamma(P_2P_3{-}kp_{4}{+}N{+}1)}
\ee
where we set $2\alpha'=1$ for simplicity. Note that
\be
P_{1,3}P_{2,4} = p_{1,3}p_{2,4} - |\vec{d}|^2  \quad {\rm while} \quad P_{1}P_{3} = p_{1}p_{3} + |\vec{d}|^2 \quad P_{2}P_{4} = p_{2}p_{4} + |\vec{d}|^2
\ee
Plugging into the (color-ordered) amplitude one finds

\be
\begin{aligned}
&{\cal A}^{\rm c.o.}_{4{+}1}[1^+, a(k), 2^-,3^+,4^-]=
 g_{op}^3 \sum_{N=0}^\infty {(r - P_3P_4 - k(p_{3}{+}p_{4}))_N \over N!} {\cal F}_N(k,p_a) {\times}\\
&\bigg[ {\varepsilon_{\alpha\beta}\ a{{\cdot}}(p_1\bar\varepsilon^{\dot\alpha\dot\beta} + k{\cdot}J_1)(P_{2}P_{3}{+}k(p_{2}{+}p_{3}){+}N{+}1) \over kp_1(P_{1}P_{4}{-}kp_3{+}N{+}1)(P_{2}P_{3}{-}kp_4{+}N{+}1)} \left(\delta_{r,1}{\delta_A^C \delta_B^D} +\delta_{r,0}{\delta_A^D \delta_B^C}\,U_{N-1,N-1}\right)\\
&-{\bar\varepsilon_{\dot\alpha\dot\beta}\ a{{\cdot}}(p_2\ \varepsilon^{\alpha\beta} + k{\cdot}J_2) \over kp_2  (P_2P_3{-}kp_{4}{+}N{+1})} \left(\delta_{r,1}{\delta_A^C \delta_B^D} +\delta_{r,0}{\delta_A^D \delta_B^C}\,V_{N-1,N-1}\right)\\&-
{\varepsilon_{\alpha\beta}\ a{{\cdot}}(p_3\bar\varepsilon^{\dot\alpha\dot\beta}  + k{\cdot}J_3)\over (P_2P_3{-}kp_{4}{+}N{+1})(P_1P_4{-}kp_{3}{+}N{+1})} \left(\delta_{r,1}{\delta_A^C \delta_B^D}  +\delta_{r,0}{\delta_A^D \delta_B^C}\,Z_{N-1,N-1}\right)\bigg]\,.
\end{aligned}
\ee
where $U_{N-1,N-1}$, $V_{N-1,N-1}$, $Z_{N-1,N-1}$ are given by
\be
\begin{aligned}
&U_{N-1,N-1}= \frac{(P_{1}P_{4}{-}kp_3{+}N{+}1)(P_{2}P_{3}{-}kp_4{+}N{+}1)}{(P_{2}P_{3}{+}k(p_{3}{+}p_{3}){+}N{+}1)(P_{2}P_{3} +N)}\\& V_{N-1,N-1}= \frac{(P_2P_3{-}kp_{4}{+}N{+1})(P_{1}P_{4} - kp_{3} + N)}{(P_{2}P_{3}{+}k(p_{3}{+}p_{3}){+}N)(P_{2}P_{3} +N)}\\&Z_{N-1,N-1}= \frac{(P_2P_3{-}kp_{4}{+}N{+1})(P_1P_4{-}kp_{3}{+}N{+1})}{(P_{2}P_{3}{+}k(p_{3}{+}p_{3}){+}N)(P_{2}P_{3} +N)}
\end{aligned}
\ee
are the required `correction factors' to ${\cal F}_N$ due to powers of $y$ and $z$, see Eq.s \,\eqref{interes} and \eqref{expan1}. While
$J_a$ are the spin generators in the spin 1/2 (L/R) representation pertaining to the (massive) gaugini of the two chiralities
\be
(J_{1})^{\mu\nu}_{\dot\alpha\dot\beta}=-{\bar \sigma}^{\mu\nu}_{\dot\alpha\dot\beta},\quad (J_{2})^{\mu\nu}_{\alpha\beta}=-{\sigma}^{\mu\nu}_{\alpha\beta} ,\quad (J_{3})^{\mu\nu}_{\dot\alpha\dot\beta}= {\bar \sigma}^{\mu\nu}_{\dot\alpha\dot\beta}\,\,.
\ee
Since ${\cal F}_N(k,p_a)\big|_{\omega=0}=1$, the soft pole ${\omega=0}$ produces the expected result that includes both the leading current terms as well as the subleading spin terms. As mentioned before, the structure of the poles is very similar to the bosonic case except for the spin terms and the absence of tachyon poles. 

\subsection{{Loops}}

As already mentioned, earlier on, one can include the effect of string loops. As a result string resonances would shift in mass and acquire a finite width. In the perturbative regime $g_s{<<}1$ these corrections would be small but at the same time would
lead to an exponential damping in $u$ (not in $R$) of the string memory \cite{ABFMhet}. Very much as in the merger of BHs or other very compact gravitating objects that leave some long-lived `remnant', the corrections due to the unstable open (super)string resonances will behave as quasi-normal modes (QNM's) with ${\rm Im}\omega \neq 0$ of the resulting unstable (non-BPS) massive open-string state. The narrower long-lived resonances would characterise and dominate the late time EM-wave signal.

While open and unoriented string amplitudes at higher loop can in principle be computed, their analysis and the systematic extraction of the necessary informations on mass-shifts and partial widths would be a formidable task that goes beyond the scope of the present investigation. 
{Here we will only make some qualitative considerations\footnote{We thank the referee for raising this issue.}. In general a macroscopic fundamental open unoriented string state can break at any point, contrary to a closed oriented string that can break only when two points meet \cite{Chialva:2003hg}. The life-time may be inferred assuming that the the decay rate is constant per unit length. This yields $\Gamma \sim g_s M$ so that ${\cal T} \sim 1/g_sM$. Notwithstanding the perturbative regime of our analysis ($g_s<<1$), typical long and macroscopic open strings may turn out to be too short-lived. However, in the Type I setting, near-BPS strings with 
$M\approx|\vec{d}|/\alpha'$, {\it i.e.} with string excitation $N<< |\vec{d}|/\sqrt{\alpha'}$, can be  made arbitrarily long-lived by tuning the parameter $|\vec{d}|$ to be large and, to some extent, $g_s$ to be (very) small. For near-BPS macroscopic strings we expect ${\cal T} \sim 1/g_sM f$, with $f= {N\sqrt{\alpha'} \over |\vec{d}| + N\sqrt{\alpha'}}$ measuring the non-BPS-ness of the state. Henceforth we will restrict our attention in the open strings with $f<<1$.}

In order for the effect that we found to give a detectable  EM-wave signal, special range of the parameters and masses, {identified above}, should be considered, whereby the usual low-energy expansion would not apply. 
At present, time resolutions that can be achieved are $\Delta{t} \approx \alpha' E \approx 10^{-15}s = 1 fs$. In turn this would  require an irrealistic value $E\approx 10^{15}GeV$ for collisions of `microscopic' open-string objects, even in the favourable $TeV$-scale (super)string scenari with $\alpha'=E_s^{-2}\approx 10^{-6} GeV^{-2}$ \cite{TeVStrings}. 

For this reason we need to consider macroscopic objects such as open-string coherent states, possibly in phenomenologically viable Type I chiral models, {satisfying the near-BPS condition $f<<1$ for their life-time to be long enough to validate our analysis}.

\section{Coherent states}
\label{Coherent}
Let us consider very massive, possibly highly charged macroscopic objects such as open cosmic {near-BPS super}strings \cite{Skliros}. A very promising approach to describe semi-classical macroscopic strings relies on the use of DDF operators \cite{DelGiudice:1971yjh} to build (open) string coherent states \cite{Skliros, MBMFcoher, AAMF}. In fact, even in less favourable scenari with Planck scale string tension $\sqrt{\alpha'}=M_{GUT}^{-1}\approx 10^{-16} GeV^{-1}$, one can construct states with large (average) mass, charge and spin. For collisions of such macroscopic objects, even for $\Delta{t} >> 1 fs$, one can find a reasonable range of masses.

In the following we will turn our attention on semi-classical coherent states. Firstly we work in the bosonic string context and then extend the analysis to the Type I superstring in the NS sector.

Let's consider coherent states in the open bosonic string. Following \cite{Skliros, MBMFcoher,AAMF} the vertex operator for a coherent state can be written as
\be
{\cal V}_{{\cal C}}(\zeta_{n},p,q;z)=\sqrt{2\alpha'} \exp \left\{ \sum_{m,n=1}^{\infty} {\zeta_{n}{\cdot}\zeta_{m}\over 2 m n} {\cal S}_{n,m} e^{-i(n{+}m)q{\cdot}X} + \sum_{n=1}^{\infty}{\zeta_{n}{\cdot}{\cal P}_{n}\over n} e^{-inq{\cdot}X} \right\}e^{ip{\cdot}X}(z)
\ee
where $q^{\mu}$ is a null momentum \emph{i.e} $q^{2}=0$, $p^{\mu}$ is a tachyonic momentum \emph{i.e}  $p^{2}=1/\alpha'$, constrained by  $2\alpha' p{\cdot}q=1$. BRST invariance is guaranteed by the fact that $\zeta_{n}{\cdot}p=\zeta_{n}{\cdot}q=0$ with polarizations of the form $\zeta_{n}^{\mu}=\lambda^{i}(\delta^{i\mu}-2\alpha'p^{i}q^{\mu})$ .

The ${\cal S}_{n,m}$ operators are symmetric in $n,m$ and read
\be
{\cal S}_{n,m}(z)=\sum_{h=1}^{n} h\, {\cal Z}_{n{-}h}({\cal U}_{\ell}^{(n)}) {\cal Z}_{m{+}h}({\cal U}_{\ell}^{(m)})
\ee
while the `generalised' momentum operators ${\cal P}^{\mu}_{n}$ are given by
\be
{\cal P}^{\mu}_{n}={1 \over \sqrt{2\alpha'}}\sum_{\ell=1}^{n} {i \partial^{\ell} X^{\mu} \over (\ell{-}1)!} {\cal Z}_{n{-}\ell}({\cal U}_{s}^{(n)})
\ee
The cycle index polynomial ${\cal Z}_{n}(u_{s})$ encode the functional dependence of both ${\cal S}_{n,m}$ and ${\cal P}^{\mu}_{n}$ on the operators ${\cal U}^{(k)}_{s}$
\be
{\cal U}_{s}^{(k)}=-i k {q{\cdot}\partial^{s}X  \over (s{-}1)!}
\ee
The simplest coherent state can be prepared as a superposition of photons. Its vertex operator can be represented as
\be
V^{ph}_{\cal{C}}=\sqrt{2\alpha'}  \exp\left({\zeta{\cdot}i\partial X \over \sqrt{2\alpha'}}\, e^{-i q{\cdot}X}+ ip{\cdot}X \right) \ee
In order to simplify the computations, the complex polarization $\zeta^{\mu}$ are chosen to satisfy $\zeta{\cdot}\zeta=0$.
Coherent states are characterized by macroscopic parameters such as the gyration radius $R_{\cal C}$, the momentum $P^{\mu}_{\cal C}$ and the total angular momentum $J^{\mu\nu}_{\cal C}$. In particular the average size of the coherent state is determined by
\be
R_{\cal C}^{2}=\langle V_{\cal C}|(X-X_{cl})^{2}| V_{\cal C} \rangle=2\alpha' \sum_{n=1}^{\infty} {|\zeta_{n}|^{2}\over n^{2}}.
\ee
The average momentum is proportional to the average level number $\langle N \rangle=\sum_{n=1}^{\infty}|\zeta_{n}|^{2} $ and reads
\be
P^{\mu}_{\cal C}=\langle V_{\cal C}|P^{\mu}| V_{\cal C} \rangle=p^{\mu}-\langle N \rangle q^{\mu}
\ee
As a consequence the average mass takes the following form
\be
M^{2}_{\cal C}=-\langle V_{\cal C}|P^{2}| V_{\cal C} \rangle={1\over \alpha'}\big(\langle N \rangle - 1\big)
\ee

Finally the total angular momentum is $J^{\mu\nu}=L^{\mu\nu}+S^{\mu\nu}$. Barring the orbital part $L^{\mu\nu}$, determined by the zero modes, the average spin is given by
\be
\langle S^{ij} \rangle=\langle V_{\cal C}|S^{ij}| V_{\cal C} \rangle={1\over \alpha'}\sum_{n=1}^{\infty}{1\over n} {Im(\lambda_{n}^{*i}\lambda_{n}^{j})}
\ee
\be
\langle S^{i-} \rangle=\langle V_{\cal C}|S^{i-}| V_{\cal C} \rangle={1\over 2\alpha' p^{+}}\sum_{n>0}\sum_{m=-\infty}^{\infty}{1\over n} {Im(\lambda_{n}^{*i} \,\lambda_{m}{\cdot}\lambda_{n{-}m})}
\ee

In order to study the EM memory effect generated by a string source where semi-classical objects are present, one can consider the scattering amplitude of two scalars $\phi$, one photon $a^{\mu}$ and a coherent state of photons
\be
{\cal A}_{3{+}1}= {{g}}^{4}_{op}{\cal C}_{D2}\prod_{j=0}^{3} \int\frac{
dz_{j}}{V_{CKG}}\left<\mathcal{V}_{\phi}(K_{1},z_{1})\mathcal{V}_{\phi}(K_{2},z_{2})\mathcal{V}_{a}(k,z_{0})\mathcal{V}^{ph}_{{\cal C}}(p_{3},q_{3},z_{3})\right>
\label{ampCohe}
\ee
Without much loss of generality, one can choose the null momentum of the coherent state $q_{3}^{\mu}$ to be collinear to the photon momentum $k^{\mu}$ \emph{i.e} $q_{3}{\cdot}k=0$. Then performing the relevant contractions, detailed in \cite{MBMFcoher}, and putting $2\alpha'=1$ one gets the following scattering amplitude\footnote{Although, as pointed out by the referee, the amplitude was derived in the context of  a single D25-brane \cite{MBMFcoher}, the result is valid also for the brane configuration under consideration at present.} 
\be
{\cal A}_{\phi \phi a {\cal C}}={g_{op}^{2}}e^{-\widehat{\zeta}{\cdot}p_{1}}{\hspace{-1mm}}\left( {a{\cdot}p_{1}\over k{\cdot}p_{1}} - {a{\cdot}p_{3}\over k{\cdot}p_{3} } - {a{\cdot}\widehat{\zeta}\,k{\cdot}p_{2}\over ( k{\cdot}p_{3}{-}1)k{\cdot}p_{3}} \right) {\Gamma(1+k{\cdot}p_{1})  \Gamma(1+k{\cdot}p_{3}) \over \Gamma(1-k{\cdot}p_{2})  }.
\ee
where $\widehat{\zeta} = \zeta e^{-iqx_0}$ and integration over $x_0$ with a factor $e^{i(p_1+p_2+p_3+k)x_0}$ is understood that implements momentum conservation level by level.
Including the other contribution, generated by the exchange $(1 \leftrightarrow 2)$, level by level, the photon profile assumes the same type of string corrections as studied in the previous sections. The crucial difference is that the coherent state can be so massive so that
$\alpha'\langle M^2\rangle = \langle N \rangle >> 1$ and the produced string correction to the memory effect would be measurable with time resolutions  $\Delta t >>1\, fs$. 
{Let us remind the reader that while keeping $ \langle N \rangle >> 1$ we need $N<<|\vec{d}|/\sqrt{\alpha'}$ in order for the coherent state to be a macroscopic excitation of a near-BPS long-lived open string. } 

The generalization of this computation to Type I superstring can be obtained using a general NS coherent vertex operator \cite{AAMF}, in the canonical super-ghost picture, restricted to the form
\be
V^{NS}_{{\cal{C}}(-1)}= \sqrt{2\alpha'}  \int d\theta \,e^{-\varphi}  \zeta_{\theta}{\cdot}\psi \, \exp\left({\zeta{\cdot}i\partial X\over \sqrt{2\alpha'}}\, e^{-i q{\cdot}X}+ ik{\cdot}X\right)
\label{susyco}
\ee
where $\zeta^{\mu}_{\theta}=\theta \zeta^{\mu} $ with $\theta$ the ${\cal N}=1$ world-sheet Grassmann variable. The Type I superstring corrections to the EM wave profile assume pretty much the same form as in the bosonic string case.

Considering a 4-pt amplitude similar to (\ref{amp1}) where, instead of using a massive doubly-charged scalar $C^{-2}$, a coherent state of the form (\ref{susyco}) is inserted one has
\be
{\cal A}_{\lambda_{+}\lambda_{+} a\, {\cal C}}= {{g}}^{4}_{op}{\cal C}_{D2}\prod_{j=0}^{3} \int\frac{
dz_{j}}{V_{CKG}}\left<\mathcal{V}_{\lambda_{+}}(K_{1},z_{1})\mathcal{V}_{\lambda^{+}}(K_{2},z_{2})\mathcal{V}_{a}(k,z_{0})\mathcal{V}^{NS}_{{\cal C}}(p_{3},q_{3},z_{3})\right>
\label{ampCoheSusy}
\ee
Choosing $\zeta_{\theta}^{M}\rightarrow{\tilde{\zeta}}_{\theta}^{j}$ to be an internal polarization, and the momentum $q_{3}^{\mu}$ collinear to $k^{\mu}$ one obtains the following result
\be
\begin{split}
{\cal A}_{\lambda_{+}\lambda_{+} a\, {\cal C}}&=g_{op}^{2}\lambda^{A}_{\alpha}(1) \lambda^{B}_{\beta}(2) \Gamma_{AB}{\cdot}\tilde{\zeta}  \,e^{-\widehat\zeta{\cdot}p_{2}} \bigg\{ {(\varepsilon^{\alpha \beta} a{\cdot}p_{2}{+} a{\cdot}\sigma^{\alpha \beta}{\cdot}k)\over k{\cdot}p_{2}} - \varepsilon^{\alpha \beta}{a{\cdot}p_{3}\over k{\cdot}p_{3}}+\\
&+ \varepsilon^{\alpha \beta} {k{\cdot}p_{1} \,a{\cdot}\widehat\zeta\over k{\cdot}p_{3}(k{\cdot}p_{3}{-}1)} \Bigg\} {\Gamma(1+k{\cdot}p_{2}) \Gamma(1+k{\cdot}p_{3})\over \Gamma(1-k{\cdot}p_{1})} + (1 \leftrightarrow 2)
\end{split}
\ee

{Recall that $\widehat{\zeta} = \zeta e^{-iqx_0}$, as in the bosonic string case, and integration over $x_0$ implementing momentum conservation is understood.} 

 The role of (minimal chiral) supersymmetry, possibly broken at a lower energy scale, {that is instrumental for the long lifetime of near-BPS long open strings, proves extremely useful in order} 
 to make the Type I model phenomenologically viable.

\subsection{{Chiral Type I models}}

So far we have considered `unrealistic' non-chiral Type I models corresponding to toroidal compactifications with D-brane separation, T-dual to Wilson line breaking \cite{Bianchi:1991eu}. 
  
Starting from the first chiral Type I model \cite{Angelantonj:1996uy}, a plethora of chiral semi-realistic models have been found and investigated. One of the general features is that open-string $U(1)$ vector bosons are anomalous and become massive via a St\"uckelberg mechanism that involves closed-string axions
\cite{U1anommass}. In Type I string model building, one of the basic phenomenological constraints is that the combination associated to weak hyper-charge  $Y=\sum_a c_a Q_a$ remain massless, before the `standard' Higgs-Englert-Brout mechanism take place \cite{YinString}.

Our analysis can be easily generalised to this case. For each $U(1)_a$ factor that contributes to $Y$ ($i.e.$ with $c_a\neq 0$) one can compute simple amplitudes  with singly charged chiral fermions (replacing the non-chiral ${\cal N}=4$ gaugini) and doubly-charged Higgs-like scalars with one or more $U(1)_a$ photon insertions. Barring the 
internal parts of the vertex operators, where $\Sigma_A$ and $\Psi^i$ would be replaced by Ramond ground-states  and NS chiral primary operators, respectively, the rest of the computation remains essentially unaltered. What matters is the non-vanishing of the Yukawa coupling, replacing $\Gamma_i^{AB}$, found in the ${\cal N}=4$ setting,  between chiral fermions and scalars\footnote{Yukawa couplings of both massless and `light' massive string states have been investigated recently in \cite{Anastasopoulos:2016yjs}.}.

In many cases the `trivial' $O(1)$ group would be replaced by some non-abelian factor, that would anyway play a marginal `spectator' role in the computations. The main difference would be the position of the poles, resulting from a mass spectrum of the form $\alpha' M^2 = n+\beta$, where  $\beta $ corresponds to some intersection angle or some internal magnetic field. The series would look slightly different but they can be summed for special kinematics.  
In any case, the `universal' nature of the string corrections to the EM (and gravitational) memory effect would not be spoiled by the `real' (mass)shift of the poles. What alters the effect significantly is an `imaginary' shift (width) due to the instability of the string resonances beyond tree-level. Assuming that $g_s$ is stabilised by fluxes or (non)perturbative effects at a small value, the leading corrections to the mass and width should scale like $\sqrt{g_s N} Q /\sqrt{\alpha'}$ where $N$ is the level and $Q$ is the charge (typically $\pm1$ or $\pm 2$ in unoriented models). There are however long-lived resonances which are narrower than the average \cite{Chialva:2003hg}. These will dominate the string corrections to the signal at late (retarded) time $u>> \alpha' n{\cdot}p$. 

One should however keep in mind that closed-string fluxes, necessary for moduli stabilisation,  and the consequent warping may change the situation quite a bit and lead to contexts similar to the ones pursued in the quest of Holographic QCD \cite{HoloQCD} or in emergent scenari for gravity, axions and `dark' photons \footnote{For recent work on holographic / emergent setups see e.g. \cite{AxionGravPhot}.}. 

\section{Discussion.}
\label{ConcOut}
We have confirmed in the Type I superstring and with higher-spin insertions that the coherent effect of the infinite tower of string resonances gives rise to a new modulated EM `memory'.\,At tree-level this string memory is oscillatory and different from the DC memory effect in gravity and in EM \cite{OldEMMemo, OldGravMemo, NewMemo}.\,It is also different  from the recently proposed `string memory effect', related to large gauge transformations of the Kalb-Ramond field \cite{Afshar1811.07368}. 
Including loops, the poles in $\omega$ would shift and broaden, producing a time-decaying signal of the same kind as for quasi-normal modes (QNMs). One should keep in mind that though fully established the effect we find can be measured with presently available detectors only if the states involved in the process are macroscopic like coherent open string states, describing for instance cosmic superstrings. 

{Notwithstanding the instability of typical macroscopic open strings, whose life-time can be estimated to be ${\cal T} \approx 1/g_sM$, we have considered near-BPS states with 
$M\approx |\vec{d}|/\alpha'$ and excitation number $1<<N<< |\vec{d}|/\sqrt{\alpha'}$, so that ${\cal T} \approx 1/g_sM f$ with $f= {N\sqrt{\alpha'}\over N\sqrt{\alpha'}+|\vec{d}|}<<1$ whose life-time can be tuned to be extremely long by proper choice of the parameter $|\vec{d}|$ and $g_s$.}

We have already argued in \cite{AAMBMFshort} that the present string memory may be related to the `global' part of the infinite (but broken) higher spin symmetries of string theory that is implemented via BRST transformations:
$\Psi \sim \Psi+{\cal Q}_{BRST} \Lambda$.  The corresponding hierarchy of `gauge symmetries' may be exposed in certain extreme regimes \cite{HigherSpinNoi, HSalter, HighEnergy}. {We  expect that a higher-spin supersymmetric version of the BMSvB symmetry, governed by some kind of (super) $W_\infty$ algebra, should exist that may realize the above physical intuition in a mathematically coherent framework. The high-energy behaviour of string scattering amplitudes remains a challenging issue that has been recently addressed in connection with chaos \cite{Gross:2021gsj}. }

 \section*{Acknowledgements}

We would like to thank A.~Addazi, D.~Consoli, G.~Di~Russo, F.~Fucito,  A.~Grillo,  {D.~Gross}, Y-t~Huang, A.~Marcian\'o,  J.~F.~Morales, {V.~Rosenhaus}, G.~C.~Rossi, D.~P.~Skliros, J.~Sonnenschein, N.~Tantalo, D.~Weissman for interesting discussions and valuable comments on the manuscript. 
The work is partly supported by the University of Tor Vergata through the Grant ID1202 ``Strong Interactions: from Lattice QCD to Strings, Branes and Holography'' within the Excellence Scheme ``Beyond the Borders'' 2019.

\end{document}